\documentclass[journal]{IEEEtran}

\usepackage{cite}

\ifCLASSINFOpdf
  \usepackage[pdftex]{graphicx}
\else
  \usepackage[dvips]{graphicx}
\fi

\usepackage{amsmath}
\usepackage{amssymb}
\usepackage{amsthm}
\makeatletter
\def\th@plain{%
  \thm@notefont{}
  \normalfont
}
\def\th@definition{%
  \thm@notefont{}
  \itshape 
}
\makeatother
\theoremstyle{definition}
\newtheorem{assume}{Assumption}

\usepackage{algorithmic}

\usepackage{array}

\ifCLASSOPTIONcompsoc
 \usepackage[caption=false,font=normalsize,labelfont=sf,textfont=sf]{subfig}
\else
 \usepackage[caption=false,font=footnotesize]{subfig}
\fi

\usepackage{url}

\def\beqa{\begin{eqnarray}}
\def\eeqa{\end{eqnarray}}
\def\beqan{\begin{eqnarray*}}
\def\eeqan{\end{eqnarray*}}

\def\C{{\mathbb{C}}}

\def\mrm{\mathrm}
\def\mbf{\mathbf}
\def\beq{\begin{equation}}
\def \eeq{\end{equation}}
\def\bbmat{\begin{bmatrix}}
\def\ebmat{\end{bmatrix}}

\def\E{\mathbb{E}}
\def\C{\mathbb{C}}

\def\epsm{\epsilon_m}
\def\phim{\phi_m}
\def\Omm{\Omega_m}

\def\omm{\omega_m}
\def\Norm{\mathcal{N}}
\def\mp{m^{\prime}}
\def\up{u^{\prime}}

\usepackage{xcolor}
\usepackage{comment}

\usepackage[hidelinks]{hyperref}

\newcommand\numberthis{\addtocounter{equation}{1}\tag{\theequation}}

\allowdisplaybreaks[2]

\begin{document}

\title{An Analytical and Experimental Study of Distributed Uplink Beamforming in the Presence of Carrier Frequency Offsets}

%
%


\author{Mehdi~Zafari,~\IEEEmembership{Graduate Student Member,~IEEE,}
        Divyanshu~Pandey,
        and~Rahman~Doost-Mohammady,~\IEEEmembership{Member,~IEEE}%
\thanks{Copyright~\copyright~2026 IEEE. Personal use of this material is permitted. However, permission to use this material for any other purposes must be obtained from the IEEE by sending a request to pubs-permissions@ieee.org.}
\thanks{Mehdi Zafari, Divyanshu Pandey, and Rahman Doost-Mohammady were with the Department of Electrical and Computer Engineering, Rice University, Houston, TX, USA (e-mail: \{mz52, dp76, doost\}@rice.edu).}%
\thanks{Mehdi Zafari is currently with the Department of Electrical Engineering and Computer Science, University of California, Irvine (UCI), CA, USA.}%
\thanks{Divyanshu Pandey is currently with Apple Inc., Sunnyvale, CA, USA.}%
\thanks{This work was supported by the U.S. National Science Foundation (NSF)
under Grants 2016727 and 2526493.}}

\maketitle

%



\begin{abstract}
Realizing distributed multi-user beamforming (D-MUBF) in time division duplex (TDD)-based multi-user MIMO (MU-MIMO) systems faces significant challenges. One of the most fundamental challenges is achieving accurate over-the-air (OTA) timing and frequency synchronization among distributed access points (APs), particularly due to residual frequency offsets caused by local oscillator (LO) drifts.
Despite decades of research on synchronization for MU-MIMO,
there are only a few experimental studies that evaluate D-MUBF techniques under imperfect frequency synchronization among distributed antennas.
This paper presents an analytical 
and experimental assessment of D-MUBF methods in the presence of frequency synchronization errors.
We provide closed-form expressions for signal-to-interference-plus-noise ratio (SINR) as a function of channel characteristics and statistical properties of carrier frequency offset (CFO) among AP antennas.
In addition, through experimental evaluations conducted with the RENEW massive MIMO testbed, we collected comprehensive datasets across various experimental scenarios.
These datasets comprise uplink pilot samples for channel and CFO estimation, in addition to uplink multi-user data intended for analyzing D-MUBF techniques.
By examining these datasets, we assess the performance of D-MUBF in the presence of CFO and compare the analytical predictions with empirical measurements.
Furthermore, we make the datasets publicly available and provide insights on utilizing them for future research endeavors.

\end{abstract}



\section{Introduction}
\label{sec:intro}

\IEEEPARstart{D}{istributed} multiple-input multiple-output (MIMO) systems~\cite{airsync, megamimo, rogalin, rashid_2022_syncAlg, Rashid_2023_syncAlg} have long been recognized for their potential to enhance coverage range, spectral efficiency, and energy efficiency of wireless networks. 
Distributed MIMO takes advantage of multiple multi-antenna access points (APs) across a shared coverage area, allowing them to jointly serve a large number of users.
For a network of distributed APs (dAPs) to efficiently serve multiple users on the same time-frequency resource, synchronization between the dAPs is essential~\cite{rogalin, martirosov_2022_syncAlg, rashid_2022_syncAlg, ahmad_2024_syncAlg, milos_2023_syncAlg}.
In infrastructure networks, utilizing high-speed, error-free optical fiber to connect dAPs can offer precise synchronization by sharing a common clock.
However, this approach may not always be practical due to the varying network topologies and the high costs of deployment.
Likewise, in ad hoc networks, such as unmanned aerial vehicles (UAVs) or sensor networks, a reliable and error-free link between distributed nodes is often not feasible. 
As a result, over-the-air (OTA) synchronization remains a crucial step for enabling distributed multi-user MIMO (MU-MIMO) transmissions.
This underscores the importance of evaluating system performance across different network topologies when employing OTA synchronization methods.

\IEEEpubidadjcol
OTA synchronization faces several challenges that significantly impact system performance~\cite{beamsync, yao_sync_stat, palat_ber_upper_bound}.
In~\cite{beamsync}, Ganesan et al. addressed the challenges of OTA carrier synchronization in distributed multi-antenna systems, emphasizing the impact of the accuracy of synchronization algorithms on beamforming performance.
In general, the accuracy of OTA synchronization methods is inherently dependent on channel conditions and the accuracy of local oscillators (LOs).
{A summary of recent developments in carrier frequency offset (CFO) estimation techniques can be found in~\cite{10702556}}.
A low signal-to-noise ratio (SNR) regime or bad channel conditions reduces the accuracy of CFO estimation, leading to inter-user interference and reduced spectral efficiency.
Therefore, residual frequency offsets, resulting either from imperfections in the synchronization process or LO drifts, are of significant concern.
The problem is further aggravated in distributed scenarios where the data transmitted or received by dAPs is affected by the combined effect of frequency offsets between all users and dAPs, making the traditional CFO compensation methods less effective. 
Thus, it is critical to characterize the impact of the uncompensated CFO among dAPs on the performance of distributed multi-user beamforming (D-MUBF).



Beamforming has been extensively studied in the context of MU-MIMO systems and distributed antenna systems, particularly with a focus on achieving optimal spectral efficiency and enhancing wireless network capacity~\cite{aryanfar_mu_bf, zafari2024asilomar, argos2012, keerthi_ul_mu_mimo}.
Most of these works do not consider non-ideal synchronization and imperfect CFO estimation and lack comprehensive experimental validations. Early studies on MU-MIMO beamforming in centralized architectures assumed ideal synchronization among APs and users \cite{argos2012}.
However, achieving accurate synchronization in distributed systems has proven to be more challenging due to the inherent independence of LOs and the distributed nature of the antennas \cite{rogalin}.
To enable distributed MU-MIMO transmission in time division duplex (TDD) mode, the practical design and implementation of synchronization techniques have been investigated for various network topologies, including wireless local area network (WLAN) \cite{rogalin}, distributed MIMO systems \cite{chorus, megamimo, airsync, airshare}, and UAV networks \cite{airbeam}.
{More recently, receiver designs have been proposed to jointly estimate CFO, channel, and data, in a MU-MIMO setting based on the orthogonal frequency division multiplexing (OFDM) by assuming Gaussian-mixture distributions for phase offsets caused by CFO~\cite{10806522}.}
However, these approaches are typically evaluated through simulations or numerical analysis under idealized conditions, and they often remain susceptible to practical impairments, such as residual CFO due to LO drifts and estimation errors, particularly in low-SNR regimes.
These imperfections can severely degrade beamforming performance in real-world deployments.

The analytical characterization of hardware impairments and synchronization errors, particularly CFO, and their impact on system performance has been extensively studied across various wireless systems, including WLAN~\cite{rogalin}, massive MIMO~\cite{Emil_Non_Ideal_HW2015, zhang2018freq_sync, ulf2014hw_impairments}, and millimeter wave (mmWave) systems~\cite{chatelier_phase_noise_mmwave}.
{For instance, a closed-form expression for SNR degradation in the presence of CFO for OFDM-based index modulation systems is derived in~\cite{besseghier2024performance}.
Although valuable, these studies do not address the impact of the statistical properties of CFO in distributed systems on beamforming.} Moreover, they lack the extensive experimental validation needed to fully understand the real-world implications of residual CFO in distributed systems.
Experimental measurements and field trials have been widely performed in several studies, including 5G-based distributed MU-MIMO~\cite{theoni2020dist_mu_mimo5g, shinya2019exp_trial5g}, 5G-based distributed MIMO in the 28 GHz band~\cite{daisuke2017outdoor_exp5g}, distributed beamforming with drones~\cite{junming2019exp_demo}, and distributed beamforming with indoor massive MIMO base station~\cite{sam2022}, among others. Most of these empirical measurements are based on a specific protocol stack or network topology and do not consider the random residual CFO in distributed systems.
To the best of our knowledge, there is a lack of experimentally validated analytical studies that specifically examine the impact of residual inter-dAP CFOs on the performance of D-MUBF.



In this paper, we fill this gap by introducing an analytical framework that studies the beamforming performance as a function of CFO statistics, thus offering a more comprehensive understanding of D-MUBF performance in the presence of random residual CFOs. We also validate our framework with experimental measurements performed using the RENEW massive MIMO testbed~\cite{renew} with perfect and imperfect synchronization.
In particular, our paper outlines three main contributions, as detailed in the following.

First, we present a straightforward, yet efficient, system model for multi-user uplink transmission based on the OFDM signaling.
We consider uncompensated residual phase offsets originating from random CFO variations between independent LOs and their effect on the channel estimation process and the calculation of beamforming weights.
We derive SINR expressions in closed form as a function of channel characteristics and statistical properties of the CFO between distributed antennas.
Our analytical model captures the impact of CFO on uplink combining in terms of inter-carrier interference (ICI) and inter-user interference (IUI), and is evaluated with different CFO characteristics through numerical results.

{Second, we conduct extensive experiments using the RENEW massive MIMO testbed~\cite{renew} in an indoor environment on the Rice University campus to evaluate D-MUBF techniques in the presence of CFO. 
We collected over 250 datasets, each containing up to 2000 frames, across various experimental scenarios.
These scenarios include MU-MIMO configurations emulating both distributed and co-located AP deployments, with and without perfect synchronization.
In the distributed scenario, each dAP in the testbed operates with an independent LO to derive carrier and sampling frequencies, resulting in random inter-dAP CFOs.
Within each dataset, every frame includes both channel sounding pilots and multi-user uplink data intended for receive beamforming.
The datasets are publicly available online~\cite{dist_mu_mimo_datasets}, and we offer insights on utilizing them for future research endeavors.}

Third, we validate the efficacy of our analytical framework using experimental datasets collected under the distributed scenario.
By comparing analytical predictions with empirical results, we offer a realistic evaluation of the practical challenges in deploying distributed MU-MIMO systems in real-world environments.
To summarize, the primary contributions of this work are as follows:

\begin{itemize}
\item We derive analytical expressions characterizing SINR degradation due to CFO in distributed beamforming scenarios, based on the statistical properties of inter-AP CFOs. This facilitates the performance evaluation of various D-MUBF techniques under realistic conditions.
\item We provide a comprehensive, publicly available dataset collected from a real-world testbed, supporting future research and validation of distributed beamforming and CFO compensation techniques.

\item We validate the analytical framework using the experimental datasets and extensive simulations across various system configurations. These evaluations offer practical insights into the selection of system parameters and CFO compensation techniques in real-world scenarios.
\end{itemize}

The remainder of the paper is organized as follows. Section II presents the system model, incorporating the effects of CFO. Section III introduces the analytical framework for deriving SINR expressions, and Section IV provides numerical and simulation results to evaluate the analysis. Section V details the experimental setup, including the testbed platform and empirical evaluation. Finally, Section VI concludes the paper.

\textbf{Notation:} Throughout the paper, scalar variables are denoted by lowercase letters (e.g., $h$), column vectors by bold lowercase letters (e.g., $\mathbf{h}$), and matrices by bold uppercase letters (e.g., $\mathbf{H}$). The sets of complex and real numbers are denoted by $\mathbb{C}$ and $\mathbb{R}$, respectively. The operators $(\cdot)^\top$ and $(\cdot)^\mathrm{H}$ represent the transpose and Hermitian (conjugate transpose), respectively. The notation $\mathcal{CN}(\mu, \sigma^2)$ denotes a circularly symmetric complex Gaussian distribution with mean $\mu$ and variance $\sigma^2$, and $\mathbb{E}\{\cdot\}$ denotes the statistical expectation. Continuous-time and discrete-time signals are represented by $f(t)$ and $f[t]$, respectively.


\section{System Model with Carrier Frequency Offsets}
\label{sec:system_model}

\begin{figure}[!t]
    \centering
    \includegraphics[width=3in]{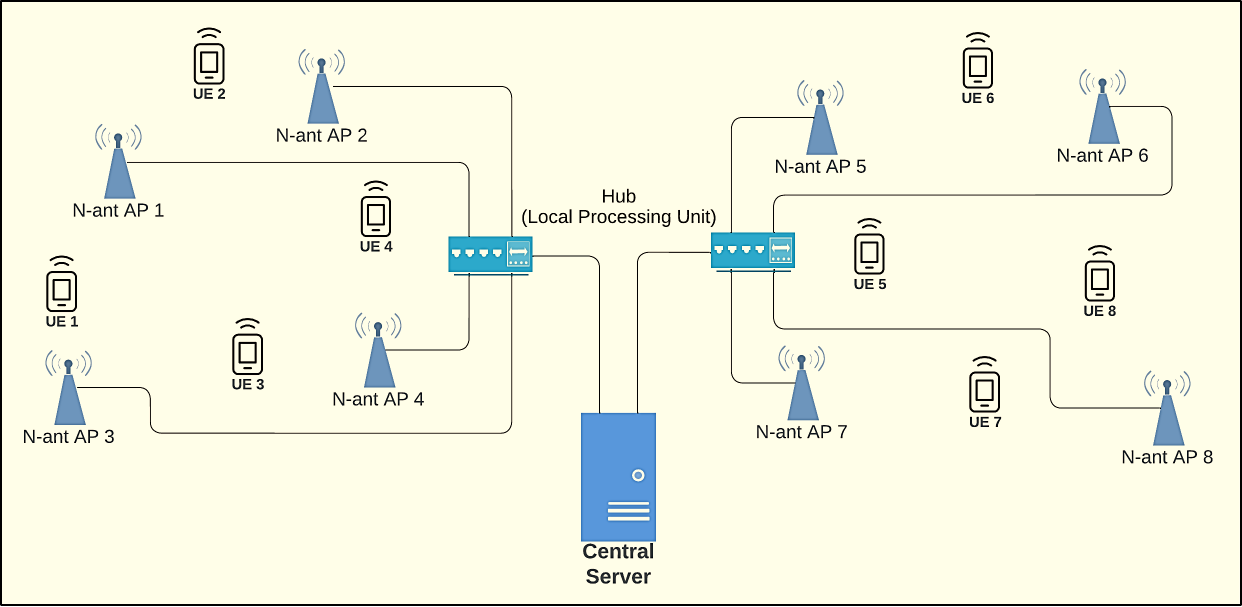}
    \caption{{Illustration of the reference network architecture, including two regions each with multiple dAPs connected to a central server (CS).}}
    \label{fig:net_arch}
\end{figure}

We consider a network of \(M\) APs with \(N\) antennas each, deployed in a given coverage area serving $K$ single-antenna users.
It is assumed that APs are connected to a central processing unit, which can be viewed as edge cloud processors~\cite{c-ran}.
To illustrate the reference network architecture, Fig.~\ref{fig:net_arch} presents an example consisting of two regions, each with a cluster of dAPs connected to a local processor and further linked to a central server (CS), jointly serving multiple users.
The CS controls cluster formation, scheduling, precoding, and beamforming operations.
The dAPs in each cluster work collaboratively to serve all the users in the coverage area.
This paper considers a fixed cluster of $M$ dAPs simultaneously serving $K$ users, without addressing the problem of cluster formation.
The transmission process follows the TDD standard, where uplink pilot, uplink data, and downlink data transmissions are separated in the time domain.

In a cluster of $M$ dAPs, for each transmission attempt, the CS determines one of the dAPs as the \textit{Master dAP}, and schedules $K$ users to be served on the same time-frequency resources.
The master dAP is responsible for broadcasting a synchronization beacon to the $K$ scheduled users. The users will then use this beacon to synchronize their frame start time with the dAPs and schedule their uplink pilot.
After transmitting orthogonal uplink pilots, all $K$ users start transmitting uplink data simultaneously.
An uplink slot comprising \(N_{\mathrm{slot}}^{\mathrm{ul}}\) OFDM symbols is considered for data transmission.
Following that, a downlink slot can be considered during which dAPs transmit precoded downlink data to the $K$ users.
For downlink transmission, a reciprocity calibration step is necessary to compensate for the non-reciprocal effects of the hardware~\cite{shepard2020argosnet}.
In this work, we only consider uplink transmission to remove the need for hardware calibration and focus exclusively on the impact of CFO on beamforming performance.

The duration of a transmission frame, containing uplink pilot and data slots for $K$ users, is denoted by $T_F$.
We consider an OFDM system with $N_{sc}$ subcarriers and total bandwidth $B$, resulting in a subcarrier spacing of $\Delta f = B / N_{sc}$.
The channel from user $k$ to antenna $i$ at AP $m$ on each subcarrier $l$ is modeled as a narrowband slow-fading channel, represented by the complex random variable $\mrm{h}_{km}^{(i)}[l] \in \mathbb{C}$, for all $l = 1, \dots, N_{sc}$.

\begin{assume}[Narrowband Slow-Fading Channel]
    \label{assume:narrowband_slow_fading}
    For each subcarrier $l = 1, \dots, N_{sc}$, the channel gain $\mrm{h}_{km}^{(i)}[l]$ is assumed to have a coherence bandwidth larger than $\Delta f$ and a coherence time longer than $T_F$. Therefore, the channel remains constant over all symbols within a transmission frame and is independently realized across different frames.
\end{assume}

The system model is designed to replicate the experimental conditions under which users are stationary and Doppler-induced phase offsets are negligible.
Considering Assumption~\ref{assume:narrowband_slow_fading}, estimated channels using uplink pilots can be used to demultiplex data symbols.
To simplify the notation, the subcarrier index $l$ is omitted where it is clear from the context.
The channel vector between user \(k\) and dAP \(m\) is denoted as \(\mathbf{h}_{km} = [\mathrm{h}_{km}^{(1)},\, \cdots\,, \mathrm{h}_{km}^{(N)}]^\top \in \mathbb{C}^{N}\).
Assuming joint coherent uplink transmission from all \(K\) users to all \(M\) dAPs, the received signal \(\mathrm{\mathbf{y}}_m^{\mathrm{ul}} \in \mathbb{C}^N\) at dAP \(m\) can be written as
\begin{equation}
    \label{eq:ul_base}
    \textstyle \mathrm{\mathbf{y}}_m^{\mathrm{ul}} = \sum_{k=1}^{K} \mathrm{\mathbf{h}}_{km}\, \mathrm{s}_k + \mathrm{\mathbf{z}}_m,
\end{equation}
where \(\mathrm{s}_k \in \mathbb{C}\) is the baseband complex-valued constellation symbol transmitted from user \(k\) with the power \(p_k\) and \(\mathrm{\mathbf{z}}_m\) is AWGN with a circularly symmetric complex Gaussian distribution, i.e. $\mathbf{z}_m \sim \mathcal{CN}\left(\mathbf{0}_N,\sigma_z^2\mathrm{\mathbf{I}}_N\right)$.
It is assumed in \eqref{eq:ul_base} that ICI is perfectly eliminated.
Based on the noise distribution, the noise samples received by each antenna are independent.
All dAPs send their received signals to the CS to be used for uplink decoding.
The collective channel vector from user \(k\) to all dAPs is \(\mathrm{\mathbf{h}}_k = \left[\mathrm{\mathbf{h}}_{k1}^\top,\, \cdots\,, \mathrm{\mathbf{h}}_{kM}^\top\right]^\top \in \mathbb{C}^{MN}\), where \(MN\) denotes the total number of service antennas used for receiving data in uplink.
In matrix notation, the channel matrix \(\mathrm{\mathbf{H}}^{\mathrm{ul}}\) is defined as \(\mathrm{\mathbf{H}}^{\mathrm{ul}} = \left[\mathrm{\mathbf{h}}_1,\, \cdots\, , \mathrm{\mathbf{h}}_K\right] \in \mathbb{C}^{MN \times K}\) containing channel gains from all users to all antennas.
The vector of uplink data samples is \(\mathrm{\mathbf{s}} = \left[\mathrm{s}_1,\,  \mathrm{s}_2,\, \cdots\,, \mathrm{s}_K\right]^\top \in \mathbb{C}^K\), and the total received signal from all dAPs can be written as
\begin{equation}
    \label{eq:ul_matrix_form}
    \mathrm{\mathbf{y}}^{\mathrm{ul}} = \mathrm{\mathbf{H}}^{\mathrm{ul}} \, \mathrm{\mathbf{s}} + \mathrm{\mathbf{z}} \ \in \mathbb{C}^{MN},
\end{equation}
where $\mathrm{\mathbf{y}}^{\mathrm{ul}} =[(\mathrm{\mathbf{y}}_1^{\mathrm{ul}})^\top, \cdots , (\mathrm{\mathbf{y}}_M^{\mathrm{ul}})^\top]^\top$, and \(\mathrm{\mathbf{z}} = [\mathrm{\mathbf{z}}_1^\top,\, \cdots\,, \mathrm{\mathbf{z}}_M^\top]^\top\) is the noise vector.


\subsection{Carrier Frequency Offset Model}
\label{subsec:cfo_model}

In a MU-MIMO setup, multiple users share the uplink slot for simultaneous data transmission.
The received signal is affected by both the user-to-dAP channel conditions and the combined impact of synchronization imperfections across each user-dAP pair.
Synchronization between a user transmitter and a dAP receiver ensures alignment in both frame timing and carrier frequency, enabling OFDM demodulation with negligible ICI.
This can be achieved using well-established synchronization schemes widely adopted in current cellular and WLAN systems~\cite{rogalin, schmidl1997}.
Conversely, in distributed MU-MIMO systems, synchronization across dAPs is critical, as each dAP operates with an independent LO with unique CFO characteristics.
Since the received signals from all cooperating dAPs are combined through receive beamforming to demultiplex user signals, precise inter-dAP synchronization is required to compensate for the relative phase rotations among dAPs.

We assume that although dAPs perform an OTA synchronization procedure, a residual relative CFO remains at each dAP.
This residual offset stems from imperfections in the OTA synchronization process, affected by channel conditions, and from LO drifts, which are influenced by hardware variability, temperature fluctuations, aging, and other environmental factors.
The residual frequency offset between dAPs disrupts subcarrier orthogonality, leading to ICI, and introduces phase shifts across successive OFDM symbols.
These phase shifts accumulate during receive combining, ultimately degrading the resulting receive SINR.

\begin{assume}[Residual Frequency Offset]
    \label{assume:freq_offset}
    The residual frequency offset at each dAP is identical across its antennas, since they share a common LO hardware, and is assumed to be constant across all subcarriers within an OFDM symbol.
\end{assume}

{Based on Assumption~\ref{assume:freq_offset}, the residual CFO at dAP $m$ can be defined as $\delta_{f,m} = f_c - f_{c,m}$,
where \(f_c\) is the true carrier frequency and \(f_{c,m}\) is the carrier frequency at the \(m\)-th dAP.}
The residual CFO $\delta_{f,m}$ can be represented as a random variable, whose statistical properties depend on the LO hardware and the accuracy of OTA synchronization.
The impact of this residual CFO in terms of ICI and phase shifts on an OFDM system is described in the following subsection.


\subsection{OFDM in the Presence of CFO}
\label{subsec:ofdm_with_cfo}

In order to provide a more comprehensible evaluation of our system model, we briefly discuss the impact of a random CFO on each subcarrier of an OFDM system.
Consider a single-user uplink transmission from user $k$ to dAP $m$ during an uplink data slot comprising $N_{\mathrm{slot}}^{\mathrm{ul}}$ OFDM symbols. Let $s[n, l]$ denote the complex baseband symbol transmitted on the $l$-th subcarrier of the $n$-th OFDM symbol within the slot.
After performing inverse fast Fourier transform (IFFT) and inserting cyclic prefix with the length of \(L_{cp}\), user \(k\) upconverts the OFDM symbols to the RF carrier \(f_c\) and transmits them through the channel.
The given dAP \(m\) receives and downconverts the signal using the carrier frequency \(f_{c,m} = f_c - \delta_{f,m}\). 
We define the normalized CFO at dAP $m$ as
\begin{equation}
    \label{eq:eps_def}
    \epsilon_m = \frac{\delta_{f,m}}{\Delta f},
\end{equation}
which is the ratio of the residual CFO to the subcarrier spacing.
At a given antenna \(i\) of dAP \(m\), the discrete-time received signal before performing the Fast Fourier Transform (FFT) can be expressed as (excluding the noise term for simplicity)
\begin{align*}
    \label{eq:ofdm_time_domain_rx_signal}
    \textstyle \mathrm{\check{y}}_m^{(i)}[n,\zeta] = \frac{1}{\sqrt{N_{sc}}}\, e^{j\frac{2\pi}{N_{sc}} (\zeta+n(N_{sc}+L_{cp})) \epsilon_m} \\ 
    \textstyle \sum_{q=0}^{N_{sc}-1} \mathrm{s}_k[n,q]\, \mathrm{h}_{km}^{(i)}[q]\, e^{j\frac{2\pi q \zeta}{N_{sc}}},
    \numberthis
\end{align*}
in which \(\zeta\) indicates the index of the time-domain OFDM samples and \(q\) represents the frequency-domain subcarrier index. 
{The frequency-domain representation of the received symbol is obtained by applying the FFT to the time-domain signal in~\eqref{eq:ofdm_time_domain_rx_signal}, with the full derivation provided in Appendix~\ref{appendix_a}.
Accordingly, the discrete-time noise-free received signal in the frequency-domain is given by
\begin{align*}
    \label{eq:ofdm_fft}
    \mathrm{y}_m^{(i)}[n,l] & \textstyle = \frac{1}{\sqrt{N_{sc}}} \sum_{\zeta=0}^{N_{sc}-1} \mathrm{\check{y}}_m^{(i)}[n,\zeta] \, e^{-j \frac{2 \pi}{N_{sc}} \zeta l} \\
    & \textstyle = e^{j\overline{\phi}_{m}[n]} \sum_{q=0}^{N_{sc}-1} \mathrm{s}_k[n,q] \, \mathrm{h}_{km}^{(i)}[q] \, G_{\epsilon}[q-l], \numberthis
\end{align*}}
where the phase rotation term \(\overline{\phi}_{m}[n]\) is defined as
\begin{equation}
    \label{eq:po_model}
    \textstyle \overline{\phi}_{m}[n] = \frac{2\pi}{N_{sc}} \left(N_{sc}+L_{cp}\right) n \epsilon_m = 2\pi \left(1+\frac{L_{cp}}{N_{sc}}\right) n\epsilon_m,
\end{equation}
{and $G_{\epsilon}[q-l]$ captures the impact of CFO in terms of ICI between subcarriers $q$ and $l$. It is a complex gain as a function of the distance between the subcarriers, i.e., $q-l$, defined as}
\begin{align*}
    \label{eq:G_def}
    G_{\epsilon}[q-l] = \frac{1}{N_{sc}} \frac{1 - e^{j 2\pi (\epsilon_m + q-l)}}{1 - e^{j\frac{2\pi}{N_{sc}} (\epsilon_m + q-l)}}, 
    \ \forall \, q = 0, \cdots, N_{sc}-1. 
    \numberthis
\end{align*}
{This gain depends on the normalized frequency offset \(\epsilon_m\) and the frequency-domain separation between the desired subcarrier \(l\) and the interfering subcarrier \(q\).}

The received OFDM symbol in~\eqref{eq:ofdm_fft} contains the desired transmitted symbol on the \(l\)-th subcarrier and ICI components due to the presence of CFO.
For the desired subcarrier $l$, i.e., when $q=l$, expression~\eqref{eq:G_def} reduces to $G_{\epsilon}[0]$, a complex gain that introduces both energy loss and a phase shift to the symbol modulated on subcarrier $l$.
Analyzing the gain $G_{\epsilon}[0]$ for a system with \(N_{sc}=64\) subcarriers demonstrates that when the CFO reaches \(5\%\) of the subcarrier spacing (e.g., \(1.5\) KHz with \(\Delta f = 30\) KHz), it results in approximately \(1\%\) energy loss per subcarrier due to ICI, along with a phase shift of \(0.05\pi\) to the received sample on that subcarrier.
Similarly, a normalized CFO of \(10\%\) causes a \(3.3\%\) energy loss and induces a \(0.1\pi\) phase shift.
To maintain per subcarrier energy loss below \(1\%\), the maximum CFO in the system must remain below \(5\%\) of the subcarrier spacing.
This underlines the impact of subcarrier spacing on the tolerance of a system to higher CFO levels.

{In the case of no CFO, i.e., $\epsilon = 0$, \eqref{eq:G_def} can be expressed as} 
\begin{align*}
    G_{0}[q-l] = \frac{1}{N_{sc}}\, \frac{1 - e^{j 2\pi (q-l)}}{1 - e^{j\frac{2\pi}{N_{sc}} (q-l)}} = \begin{cases}
        1 & q = l \\
        0 & q \neq l
    \end{cases},
    \numberthis
\end{align*}
which indicates perfect orthogonality between subcarriers.
{For evaluating $G_0 [q-l]$ at $q=l$, L'Hôpital's rule can be used.}
Besides the ICI term in \eqref{eq:ofdm_fft}, there is also the phase rotation term \eqref{eq:po_model}, which depends on both the symbol index \(n\) and the CFO term \(\epsilon_m\), and remains constant across all subcarriers of a symbol. 
{Accordingly, \eqref{eq:ofdm_fft} can be further written as
\begin{align*}
    \label{eq:ofdm_fft_simplified}
    \mathrm{y}_m^{(i)}[n,l] & \textstyle = e^{j\overline\phi_{m}[n]}\, \mathrm{s}_k[n,l]\, \mathrm{h}_{km}^{(i)}[q]\, G_{\epsilon}[0] \\
    & \textstyle \quad + e^{j\overline\phi_{m}[n]}\, \sum_{q=0, q \neq l}^{N_{sc}-1} \mathrm{s}_k[n,q]\, \mathrm{h}_{km}^{(i)}[q]\, G_{\epsilon}[q-l] \\
    & \textstyle = \mathrm{s}_k[n,l] \, \mathrm{h}_{km}^{(i)}[l]\, G_{\epsilon}[0]\, e^{j\overline\phi_{m}[n]} + \mathrm{ICI}_{\epsilon}[l],
    \numberthis
\end{align*}
which includes the desired sample affected by the CFO in addition to the interference from other subcarriers.}
It is straightforward to verify that when \(\epsilon = 0\), the noise-free received symbol at subcarrier \(l\) simplifies to \(\mathrm{y}_m^{(i)}[n,l] = \mathrm{s}_k[n,l]\, \mathrm{h}_{km}^{(i)}[l]\), indicating that the symbol is scaled solely by the narrowband fading channel on that subcarrier, without any ICI.

The ICI power can be quantified based on the channel gains and the complex term defined in~\eqref{eq:G_def}.
The dominant interference on a given subcarrier $l$ originates from its adjacent subcarriers, i.e., \(q = l \pm 1\), with the complex gain \(G_{\epsilon}[1]\).
For low ranges of CFO, e.g., \(|\epsilon_m| \leq 5\%\), the largest energy gain of interfering subcarriers is \(\left| G_{\epsilon}[1] \right|^2 < 2.3\times 10^{-3}\) ($-26.4$ dB).
Consequently, the additive interference term \(\mathrm{ICI}_{\epsilon}[l]\) in \eqref{eq:ofdm_fft_simplified} has negligible energy, i.e., \(\left| \mathrm{ICI}_{\epsilon}[l] \right|^2 \ll 1\).
Therefore, it can be treated as an additional contribution to the effective noise on each subcarrier, while the direct impact of CFO on the desired subcarrier is captured by the gain $G_{\epsilon}[0]$.

Based on the above analysis, we derive a simplified OFDM uplink signal model that incorporates the effects of CFO as
\begin{align*}
    \label{eq:ofdm_signal_model_su}
    \mathrm{y}_m^{(i)}[n,l] 
    = \mathrm{s}_k[n,l] \, \mathrm{h}_{km}^{(i)}[l]\, \Omega_{m}[n] + \mathrm{z}_m[n,l],
    \numberthis
\end{align*}
where \(\mathrm{z}_m[n,l]\) is the overall additive noise and \(\Omega_{m}[n]\) represents the overall energy loss and phase rotation on each subcarrier.
Based on \eqref{eq:ofdm_fft_simplified}, the complex gain \(\Omega_{m}[n]\) for a given dAP \(m\) and OFDM symbol index \(n\) is expressed as
\begin{align*}
    \label{eq:omega_def}
    \Omega_{m}[n] 
    &= \frac{1}{N_{sc}} \, \frac{\sin(\pi \epsilon_m)}{\sin(\frac{\pi \epsilon_m}{N_{sc}})} \, e^{j\left[2 n \left(1 + \frac{L_{cp}}{N_{sc}}\right) + \frac{N_{sc}-1}{N_{sc}}\right] \pi \epsilon_m}, \numberthis
\end{align*}
and we define $\Omega_{m}[n] \triangleq \omega_{m} \, e^{j \phi_{m}[n]}$, which is composed of a magnitude \(|\Omega_{m}[n]| = \omega_{m}\) and a phase \(\angle \Omega_{m}[n] = \phi_{m}[n]\).
The term $\phi_{m}[n]$ represents the overall phase shift on the \(n\)-th OFDM symbol due to the CFO.
Finally, using the single-user representation in \eqref{eq:ofdm_signal_model_su}, the uplink system model for multi-user transmission in the presence of a residual CFO, for a fixed subcarrier \(l\) and OFDM symbol \(n\), is expressed as
\begin{equation}
    \label{eq:ul_sytem_model}
    \textstyle \mbf{y}_m^{\mrm{ul}} = \sum_{k = 1}^K \Omega_m \, \mbf{h}_{km} \, \mrm{s}_k + \mbf{z}_m,
\end{equation}
where \(\mbf{y}_m^{\mrm{ul}} \in \C^{N}\) demonstrates the vector of received uplink signals at the \(m\)-th dAP, and \(\Omega_m\) is defined in \eqref{eq:omega_def} for a given OFDM symbol index \(n\). We dropped the subcarrier and OFDM symbol indices for brevity of notation.


\section{Analytical Framework for D-MUBF}
\label{sec:analytical}

After transmitting orthogonal uplink pilots, users will start transmitting their data simultaneously.
Using the received uplink pilots, the channel vectors from users to dAPs and the user-dAP CFO are estimated.
To mitigate inter-user interference, multi-user receive beamforming is employed, where beamforming vectors are calculated based on the channel estimates.
As described in Assumption~\ref{assume:freq_offset}, each dAP uses an independent high-stability reference LO to generate a common reference signal for all antennas.
However, uncompensated CFO between dAPs can substantially degrade distributed beamforming performance.
In this section, we derive closed-form SINR expressions for D-MUBF as a function of channel vectors and statistical properties of the inter-dAP residual CFO component, defined in \eqref{eq:omega_def}.


\subsection{Uplink Multi-User Beamforming}
\label{subsec:uplink_bf}

Using the combining weight vector \(\mathbf{w}_{km} \in \C^{N}\), designed to detect the signal from user \(k\), the estimated received symbol can be obtained from~\eqref{eq:ul_sytem_model} as
\begin{alignat*}{2}
    \label{eq:ul_est_sample}
    \widehat{\mathrm{s}}_{k} = &\textstyle \sum\limits_{m=1}^{M} \mathbf{w}_{km}^{\mathrm{H}}\, \mathbf{y}_{m}^{\mathrm{ul}} = \sum\limits_{m=1}^{M} \Omega_{m} \, \mathbf{w}_{km}^{\mathrm{H}} \, \mathbf{h}_{km}\, \mathrm{s}_k \\
    &\textstyle + \sum\limits_{m=1}^{M} \Omega_m \, \mathbf{w}_{km}^{\mathrm{H}} \, \sum\limits_{u=1, u\neq k}^{K} \mathbf{h}_{um} \mathrm{s}_u + \sum\limits_{m=1}^{M} \mathbf{w}_{km}^{\mathrm{H}} \mathbf{z}_m.
    \numberthis
\end{alignat*}
It is considered in this analysis that all \(N\) antennas at all \(M\) dAPs are participating in beamforming.
Previous simulations presented in~\cite{sam2022} indicated that in distributed uplink beamforming, where there is no perfect frequency synchronization between dAPs, increasing the number of receive antennas in beamforming does not necessarily lead to higher received SNR or reduced error vector magnitude (EVM).
{Notably, the experimental results in~\cite{sam2022}, evaluating receive beamforming using zero-forcing and conjugate methods (Figs. 4 and 5, respectively), demonstrate that in the presence of CFO, increasing the number of beamforming antennas from 8 to 64 provides no improvement in EVM for the single-user scenario and only limited improvement for multi-user scenarios.}

Using the estimated signals in~\eqref{eq:ul_est_sample}, the average uplink SINR for a given user \(k\) can be derived when all \(M\) dAPs are receiving the uplink signals and cooperating in beamforming.
By considering the channel vectors from different users uncorrelated, e.g., due to enough spatial separation and rich scattering, and by defining the transmit SNR of each user as \(\gamma_k = p_k / \sigma^2_z\), the average uplink SINR can be expressed as
\begin{align*}
    \label{eq:sinr_simplified}
    & \mathrm{SINR}_k = \numberthis \\
    & \frac{\gamma_k\, \mathbb{E}\{|\sum_{m=1}^{M} \mathbf{w}_{km}^{\mathrm{H}} \, \Omega_{m} \, \mathbf{h}_{km} |^2\}}{\sum\limits_{u\neq k}^{K} \gamma_u\, \mathbb{E}\{|\sum_{m=1}^{M} \mathbf{w}_{km}^{\mathrm{H}} \, \Omega_{m} \, \mathbf{h}_{um} |^2\} + \sum\limits_{m=1}^{M} \mathbb{E}\{\|\mathbf{w}_{km}^{\mathrm{H}} \|^2\}}, 
\end{align*}
in which the expectations are with respect to the joint probability distribution of the channel vectors and the CFO component.

In D-MUBF methods, beamforming weights are derived based on channel estimates from users to mitigate inter-user interference~\cite{zafari2024asilomar}.
The performance of the D-MUBF methods depends on several factors, including number of users, number of antennas at each dAP, spatial correlation between antennas, and residual phase and frequency offsets~\cite{aryanfar_mu_bf}.
{D-MUBF can be implemented using various beamforming techniques. In this work, we focus on two widely used approaches: conjugate (CBF)~\cite{giovanni2021cbf} and zero-forcing beamforming (ZFBF)~\cite{hong2025zfbf}. These methods are selected due to their practical relevance and complementary characteristics—CBF is favored for its simplicity and robustness to channel estimation errors, while ZFBF provides higher interference suppression at the cost of increased computational complexity. Evaluating both allows us to assess D-MUBF performance across a representative trade-off space between implementation complexity and interference mitigation.}
To better understand the impact of CFO on SINR, we derive a conditional closed-form expression for the SINR, evaluated for fixed channel vectors and expressed as a function of the statistical properties of the inter-dAP CFO component.
This enables the evaluation of SINR with respect to the inter-dAP CFO distribution for any given channel realization.


\subsection{Closed-Form Expression for SINR}
\label{subsec:closed_form_sinr}

In general, computing the average SINR in closed form requires the joint probability distribution of the channel vectors and CFO.
However, in this analysis, we fix the channel vectors and focus exclusively on the impact of CFO.
We first characterize the distribution and statistical properties of the complex CFO term defined in~\eqref{eq:omega_def}, and then derive a conditional SINR expression as a function of these CFO statistics.

\subsubsection{Distribution and Moments of CFO Component}
\label{subsubsec:dist_exp_omega}

The inter-dAP CFO component \eqref{eq:omega_def} for a fixed OFDM symbol index \(n\) can be shown as \(\Omega_m = \omega_m \, e^{j \phi_m}, \; \forall m = 1, \cdots, M\).
Both the magnitude and the phase of \(\Omega_m\) are functions of the normalized CFO defined in~\eqref{eq:eps_def} for every dAP \(m\).
We model the normalized CFO $\epsilon_m$ as a random variable that is statistically independent of the channel.
The distribution of $\epsilon_m$ can be empirically estimated for a given platform.
In this analysis, we consider two representative distributions—normal and uniform—to evaluate the impact of different CFO statistics on system performance.
In both cases, the random variable $\phi_m$ follows a scaled version of the distribution of $\epsilon_m$, since $\phi_m$ is a linear function of $\epsilon_m$.
However, $\omm$ is a nonlinear function of $\epsilon_m$, making its statistical analysis more challenging.
To address this, we introduce an approximation for $\omm$ that enables the derivation of its first- and second-order moments.

Since in this analysis we are interested in small ranges of CFO, i.e., \(|\epsm| \leq 0.1 \, (10\%)\), we can use the third-order Taylor approximation for the $\sin(\cdot)$ functions in~\eqref{eq:omega_def} as
\begin{align*}
    \omega_{m} 
    &\approx \frac{1}{N_{sc}} \, \frac{\pi \epsm - \frac{(\pi \epsm)^3}{6}}{\frac{\pi \epsilon_m}{N_{sc}} - \frac{(\pi \epsilon_m / N_{sc})^3}{6}} = \frac{1 - \frac{(\pi \epsm)^2}{6}}{1 - \frac{(\pi \epsm)^2}{6 N_{sc}^2}}.
    \numberthis
\end{align*}
Given that we focus on small values of $\epsilon_m$, and that the number of subcarriers $N_{sc}$ is typically large (e.g., $N_{sc} \geq 8$), we apply the approximation $(1 - x)^{-1} \approx 1 + x$ for $x \ll 1$ to simplify the expression as
\begin{equation}
    \textstyle \omega_{m} \approx 1 - \frac{(\pi \epsm)^2}{6}, \quad \text{for} \; |\epsm| < 0.4, \; N_{sc} \geq 8,
    \label{eq:omm_approx}
\end{equation}
which is a quadratic function of \(\epsm\).
Using this approximation, the final form of the CFO component is
\begin{equation}
    \label{eq:Omega_approax}
    \textstyle \Omm = \omm \, e^{j \phim} \approx \left(1 - \frac{\pi^2  \epsm^2}{6}\right) \, e^{j g_n \pi \epsm}
\end{equation}
where, for notational simplicity, we define $g_n$ to represent the following term
\begin{equation}
    \textstyle g_n \triangleq \left[2 n \left(1 + \frac{L_{cp}}{N_{sc}}\right) + \frac{N_{sc}-1}{N_{sc}}\right].
\end{equation}

Using the approximation in~\eqref{eq:Omega_approax}, the first- and second-order moments of \(\Omm\) can be derived for any specific distribution of \(\epsm\). {First, let \(\epsm\) be a zero-mean uniformly distributed random variable as $\epsm \sim \mathcal{U} [-\alpha, \alpha]$.}
Note that the distribution parameter \(\alpha\) does not depend on the dAP index \(m\) since \(\epsm\) for \(m=1, \cdots, M\) are represented as independent and identically distributed (i.i.d.) random variables.
Thus, we drop the subscript \(m\) and evaluate the expectation terms \(\E\{\Omega\}\) and \(\E\{|\Omega|^2\}\).
For the uniform distribution, the expectations are 
\begin{align*}
    & \textstyle \E\{\Omega\} 
    = \int_{-\alpha}^\alpha \left(1 - \frac{\pi^2 \, \epsilon^2}{6}\right) \, e^{j g_n \pi \epsilon} \, \frac{1}{2\alpha} \, d\epsilon
    \label{eq:exp_omega_uniform} \numberthis \\
    & \textstyle = \frac{1}{g_n} \, \left[\frac{1}{\pi \alpha} - \frac{\pi \alpha}{6} + \frac{1}{3 \pi \alpha g_n^2}\right] \, \sin(g_n \pi \alpha) - \frac{1}{3 g_n^2} \, \cos(g_n \pi \alpha),  \\
    & \textstyle \E\{|\Omega|^2\} = \frac{\pi^4 \, \alpha^4}{180} - \frac{\pi^2 \, \alpha^2}{9} + 1. \label{eq:exp_omega_sq_uniform} \numberthis
\end{align*}
{For a normal distribution as $\epsm \sim \Norm(0, \beta^2)$,}
where \(\beta^2\) is the variance of \(\epsm\), the expectation terms are derived as
\begin{align*}
    & \textstyle \E\{\Omega\} = \left[1 - \frac{\pi^2 \beta^2}{6} \left(1 - g_n^2 \pi^2 \beta^2\right)\right] \, e^{- \frac{1}{2} g_n^2 \pi^2 \beta^2}, \label{eq:exp_omega_normal} \numberthis \\
    & \textstyle \E\{|\Omega|^2\} = \frac{\pi^4 \, \beta^4}{12} - \frac{\pi^2 \, \beta^2}{3} + 1. \label{eq:exp_omega_sq_normal} \numberthis
\end{align*}
Next, we derive a conditional expression for the SINR, leveraging the closed-form first- and second-order moments of the CFO term for both the normal and uniform distributions.

\subsubsection{SINR Expression with Fixed Channel}
\label{subsubsec:cond_sinr}

We derive the conditional variances of the desired signal, interference, and noise term to obtain SINR for a known channel.
The estimated modulated symbol using our system model is given in \eqref{eq:ul_est_sample}.
Given that all channel vectors are known, we can define the SINR as a function of \(\mbf{H}^{\mrm{ul}}\)  from \eqref{eq:ul_matrix_form} as
\begin{equation}
    \label{eq:cond_sinr_main}
    \text{SINR}(\mbf{\mbf{H}^{\mrm{ul}}}) = \frac{P_{s|\mbf{\mbf{H}^{\mrm{ul}}}}}{P_{v|\mbf{\mbf{H}^{\mrm{ul}}}} + P_{z|\mbf{\mbf{H}^{\mrm{ul}}}}},
\end{equation}
where \(P_{s|\mbf{\mbf{H}^{\mrm{ul}}}}\), \(P_{v|\mbf{\mbf{H}^{\mrm{ul}}}}\), and \(P_{z|\mbf{\mbf{H}^{\mrm{ul}}}}\) denote the conditional powers of the desired signal, interference, and noise, respectively. 
{The detailed derivation of the conditional desired signal power is provided in Appendix~\ref{appendix_b}, resulting in}
\begin{align*}
    P_{s|\mbf{\mbf{H}^{\mrm{ul}}}}
    & \textstyle = p_{\mrm{ul}} \, \sum\limits_{m=1}^{M} \E\{|\Omega|^2\}\, |\mathbf{w}_{km}^{\mathrm{H}} \mathbf{h}_{km}|^2 \numberthis \\
    & \textstyle \quad + p_{\mrm{ul}} \, \sum\limits_{m=1}^M \sum\limits_{\mp \neq m}^M \left(\E\{\Omega\}\right)^2 \, \left(\mbf{w}_{km}^{\mrm{H}} \mbf{h}_{km} \mbf{h}_{k\mp}^{\mrm{H}} \mbf{w}_{k\mp}\right),
\end{align*}
where $p_{\mathrm{ul}}$ denotes the uplink transmit power, which is assumed to be equal across all users for simplicity.
The conditional power of the inter-user interference term can be written as
\begin{align*}
    P_{v|\mbf{\mbf{H}^{\mrm{ul}}}} = \textstyle p_{\mrm{ul}} \, \sum\limits_{m=1}^{M} \E\{|\Omega|^2\}\, \sum\limits_{u\neq k}^K \sum\limits_{\up \neq k}^K \mathbf{w}_{km}^{\mathrm{H}} \mathbf{h}_{um} \mathbf{h}_{\up m}^{\mathrm{H}} \mathbf{w}_{km} \\
    \textstyle +\, p_{\mrm{ul}} \, \sum\limits_{m=1}^M \sum\limits_{\mp \neq m} \left(\E\{\Omega\}\right)^2 \sum\limits_{u\neq k}^K \sum\limits_{\up \neq k}^K \mathbf{w}_{km}^{\mathrm{H}} \mathbf{h}_{um} \mathbf{h}_{\up \mp}^{\mathrm{H}} \mathbf{w}_{k\mp},
    \numberthis
\end{align*}
{with the detailed derivation provided in Appendix~\ref{appendix_b}.}
The last term is the conditional noise power as
\begin{align*}
    P_{z|\mbf{\mbf{H}^{\mrm{ul}}}} = \textstyle 
    \sum\limits_{m=1}^{M} \E\{| \mathbf{w}_{km}^{\mathrm{H}} \, \mathbf{z}_m |^2 \, | \, \mbf{H}^{\mrm{ul}}\} = \sigma_z^2 \, \sum\limits_{m=1}^{M} \|\mathbf{w}_{km}\|^2,
    \numberthis
\end{align*}
given that the noise is distributed as $\mathbf{z}_m \sim \mathcal{CN}\left(\mathbf{0}_N,\sigma_z^2\mathrm{\mathbf{I}}_N\right)$.
{By substituting the conditional power terms into~\eqref{eq:cond_sinr_main}, we obtain the closed-form SINR expression shown in~\eqref{eq:sinr_closed_form}, which appears at the top of the next page.}
The expectation terms in~\eqref{eq:sinr_closed_form} are with respect to the CFO distribution and can be calculated if the distribution is known.
For uniform and normal distributions of CFO, we derived these statistics in \eqref{eq:exp_omega_uniform}, \eqref{eq:exp_omega_sq_uniform}, \eqref{eq:exp_omega_normal}, and \eqref{eq:exp_omega_sq_normal}, respectively.
Given that the channel vectors are known, the beamforming weight vectors $\mathbf{w}_{km}, \forall k, m,$ are also known, as they are functions of the columns of $\mathbf{H}^{\mathrm{ul}}$.
In the following section, we perform numerical and simulation analysis to evaluate the SINR expression in~\eqref{eq:sinr_closed_form}.

\begin{figure*}[t]
\normalsize
\begin{multline*}
    \label{eq:sinr_closed_form}
    \text{SINR}\left(\mbf{H}^{\mrm{ul}}\right) = \\
    \frac{\E\{|\Omega|^2\}\, \sum\limits_{m=1}^{M} |\mathbf{w}_{km}^{\mathrm{H}} \mathbf{h}_{km}|^2 + \left(\E\{\Omega\}\right)^2 \, \sum\limits_{m=1}^M \sum\limits_{\mp \neq m}^M \mbf{w}_{km}^{\mrm{H}} \, \mbf{h}_{km} \, \mbf{h}_{k\mp}^{\mrm{H}} \,  \mbf{w}_{k\mp}}{\E\{|\Omega|^2\}\, \sum\limits_{m=1}^{M} \sum\limits_{u\neq k}^K \sum\limits_{\up \neq k}^K \mathbf{w}_{km}^{\mathrm{H}} \, \mathbf{h}_{um} \, \mathbf{h}_{\up m}^{\mathrm{H}} \, \mathbf{w}_{km} + \left(\E\{\Omega\}\right)^2 \, \sum\limits_{m=1}^M \sum\limits_{\mp \neq m}^M \sum\limits_{u\neq k}^K \sum\limits_{\up \neq k}^K \mathbf{w}_{km}^{\mathrm{H}} \, \mathbf{h}_{um} \, \mathbf{h}_{\up \mp}^{\mathrm{H}} \, \mathbf{w}_{k\mp} + \frac{\sigma_z^2}{p_{\mrm{ul}}} \, \sum\limits_{m=1}^{M} \|\mathbf{w}_{km}\|^2}
    \numberthis
\end{multline*}
\hrulefill
\end{figure*}


\section{Numerical and Simulation Results}
\label{sec:num_results}

In this section, we present numerical and simulation results to evaluate the closed-form SINR expression in~\eqref{eq:sinr_closed_form} as a function of the channel realization and the statistical properties of the inter-dAP CFO.


\subsection{Numerical Evaluation of the SINR Expression}
\label{subsec:eval_sinr_exp}

The SINR expression in~\eqref{eq:sinr_closed_form} is a function of channel vectors and the expectations of the CFO component in the forms of \(\E\{\Omega\}\) and \(\E\{|\Omega|^2\}\).
We evaluate the asymptotic behavior of the SINR with respect to the standard deviation (STD) of the normalized CFO component, defined in~\eqref{eq:eps_def}.
{The expectation term $\E\{\Omega\}$ for both CFO distributions, uniform (given in~\eqref{eq:exp_omega_uniform}) and normal (given in~\eqref{eq:exp_omega_normal}), decays and converges to zero as the STD increases (i.e., $\alpha/\sqrt{3}$ for uniform and $\beta$ for normal), and also with increasing OFDM symbol index $n$ in the time domain.
In addition, from~\eqref{eq:exp_omega_sq_uniform} and~\eqref{eq:exp_omega_sq_normal}, it can be inferred that $\left(\E\{\Omega\}\right)^2$ approaches zero when $n \alpha/\sqrt{3} > 0.25$ for the uniform distribution and $n \beta > 0.25$ for the normal distribution.}
Therefore, within these specified ranges of $n$, $\alpha$, and $\beta$, the term $\left(\E\{\Omega\}\right)^2$ can be asymptotically excluded from the SINR expression in~\eqref{eq:sinr_closed_form}.
We note that the distribution parameters of the normalized CFO term in~\eqref{eq:eps_def} are chosen to ensure the validity of the approximation introduced in~\eqref{eq:omm_approx}.

\begin{figure}[!t]
\centering
\subfloat [{Normal Distribution of CFO: $\epsilon \sim \mathcal{N}(0, \beta^2)$}] {
    \includegraphics[width=0.9\linewidth]{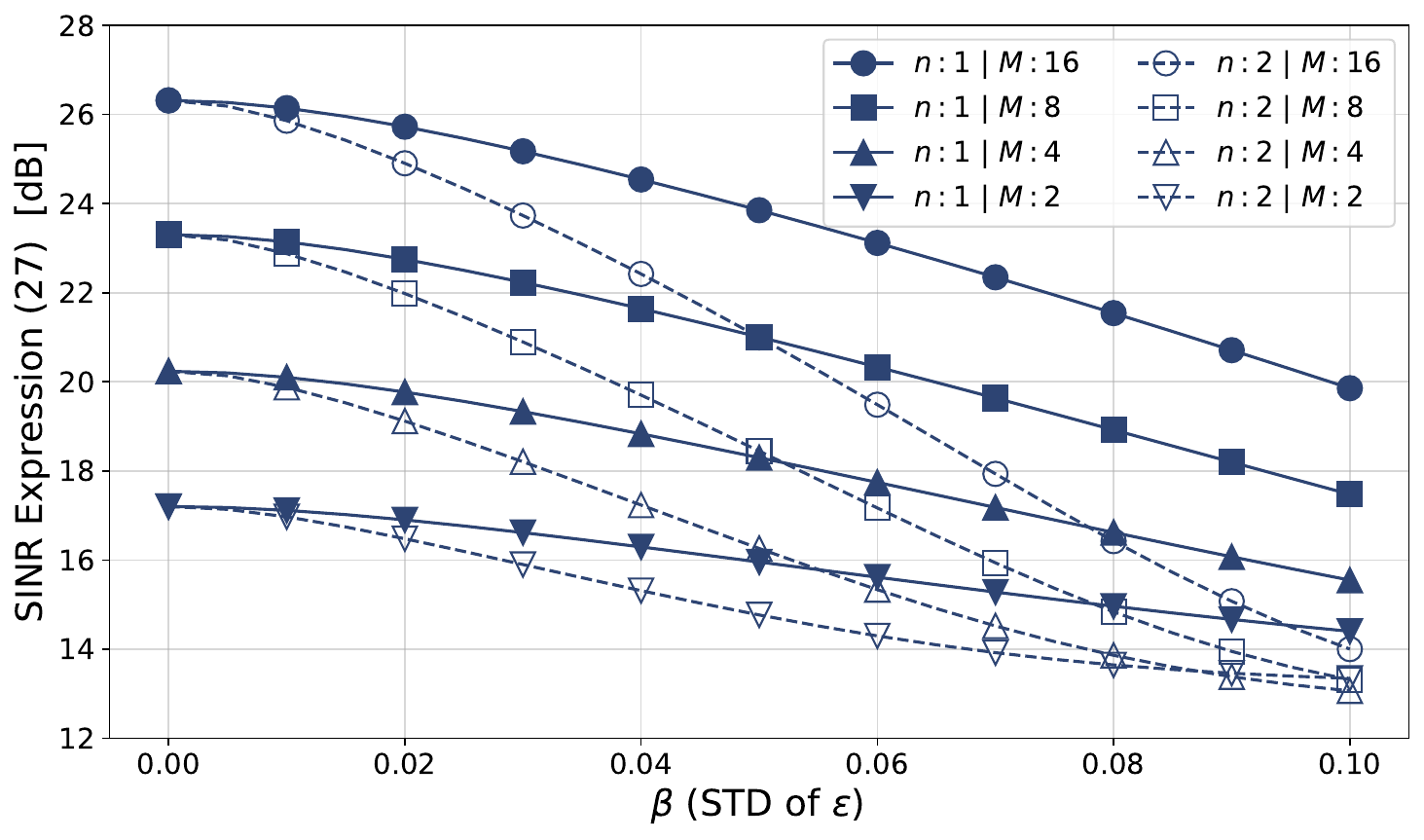}
    \label{fig:sinr_std_normal}
    }
    
\subfloat [{Uniform Distribution of CFO: $\epsilon \sim \mathcal{U}{[-\alpha, \alpha]}$}] {
    \includegraphics[width=0.9\linewidth]{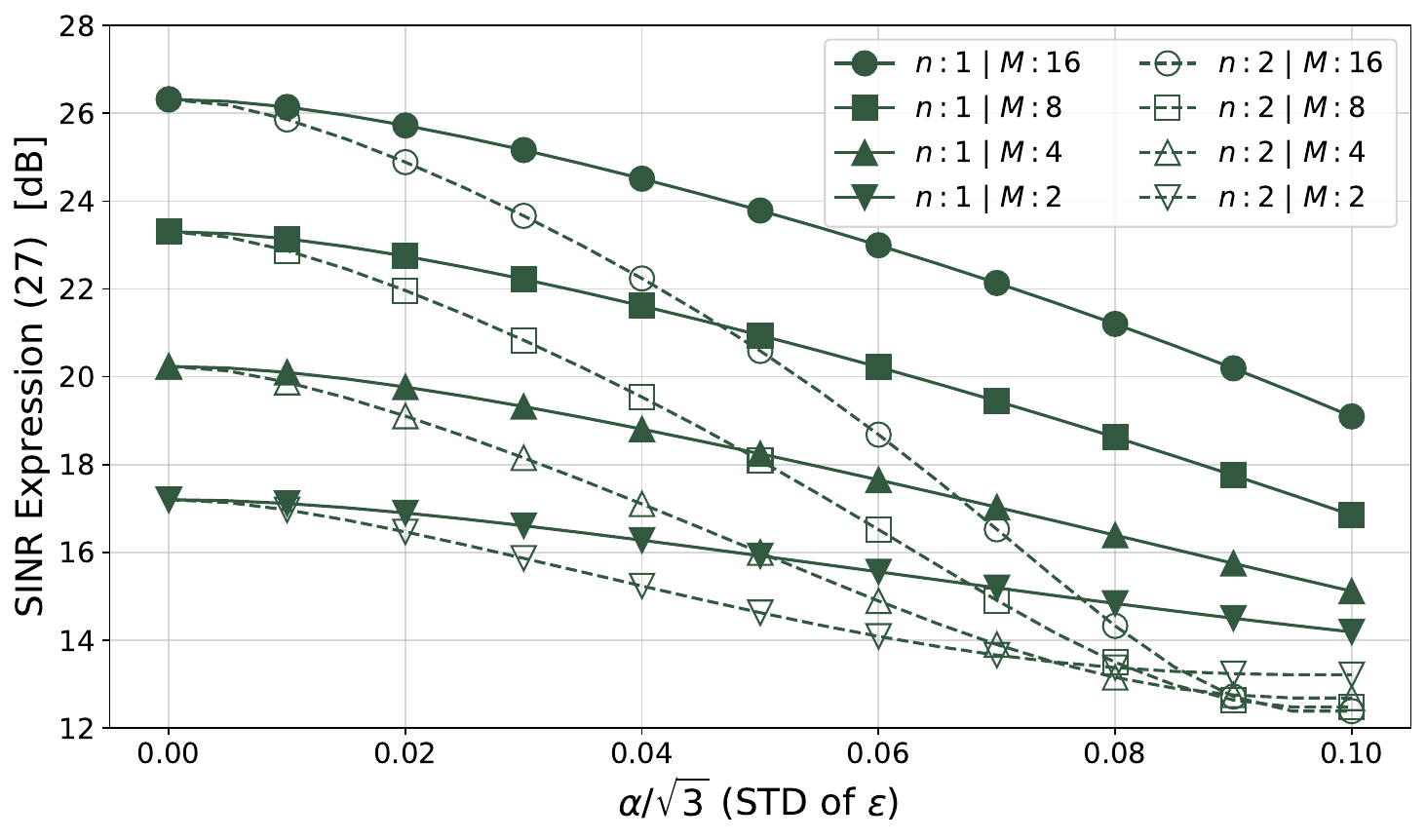}
    \label{fig:sinr_std_uniform}
    }
\caption{{The conditional SINR expression in~\eqref{eq:sinr_closed_form} versus STD of the normalized CFO in~\eqref{eq:eps_def}, distributed as a \textbf{(a)} normal and \textbf{(b)} uniform random variable. There are \(N=16\) antennas per dAP and \(K=2\) users, and CBF is applied.}}
\label{fig:sinr_std_M}
\end{figure}

The asymptotic behavior of the conditional SINR with respect to the STD of CFO can be expressed as,
\begin{align*}
    \label{eq:sinr_asymptotic}
    &\text{SINR}(\mbf{H}^{\mrm{ul}}) \approx \\
    &\frac{\sum\limits_{m=1}^{M} |\mathbf{w}_{km}^{\mathrm{H}} \mathbf{h}_{km}|^2}{\sum\limits_{m=1}^{M} \sum\limits_{u\neq k} \sum\limits_{\up \neq k} \mathbf{w}_{km}^{\mathrm{H}} \mathbf{h}_{um} \mathbf{h}_{\up m}^{\mathrm{H}} \mathbf{w}_{km} + \frac{1}{\E\{|\Omega|^2\}} \frac{\sigma_z^2}{p_{\mrm{ul}}} \sum\limits_{m=1}^{M} \|\mathbf{w}_{km}\|^2}, \\
    &\forall \, n\alpha/\sqrt{3} > 0.25 \; \text{(uniform)} \; \text{or} \; \forall \, n \beta > 0.25 \; \; \text{(normal)}, \numberthis
\end{align*}
which shows that the only dependency on the CFO distribution that remains in the asymptotic SINR is the term \(\E\{|\Omega|^2\}\), derived in~\eqref{eq:exp_omega_sq_uniform} and \eqref{eq:exp_omega_sq_normal} for uniform and normal distributions, respectively.
To assess the SINR expression with both distributions, we generated numerical results based on the conditional expression in~\eqref{eq:sinr_closed_form} for CBF with Rayleigh fading channel.
The results are illustrated in Fig.~\ref{fig:sinr_std_M} for different numbers of dAPs (\(M\)) and two values of the OFDM symbol index (\(n\)).

The SINR values in Fig.~\ref{fig:sinr_std_M} demonstrate that by doubling the number of dAPs in the low-CFO regime, the SINR increases by approximately \(3\) dB.
As the STD of the random inter-dAP CFO term increases, the SINR drops until a point where it becomes almost independent of the CFO distribution parameter.
This is because our numerical results apply the approximation \eqref{eq:omm_approx}, which allows us to utilize \eqref{eq:exp_omega_uniform}--\eqref{eq:exp_omega_sq_normal} for representing the CFO component.
Consequently, at higher STD levels, the SINR converges to the asymptotic expression in~\eqref{eq:sinr_asymptotic}, showing minimal dependence on the statistical properties of the CFO.
Fig.~\ref{fig:sinr_std_M} further indicates that the choice between normal and uniform CFO distributions has negligible impact on the resulting SINR.
Hence, for the remainder of our analysis, the normal CFO distribution is used as $\epsilon \sim \mathcal{N}(0, \beta^2)$.
In the next subsection, rather than numerically evaluating~\eqref{eq:sinr_closed_form}, we simulate an OFDM-based uplink transmission process, introducing CFO as random realizations drawn from the normal distribution defined above. We then measure the received signal power to compute the SINR and compare the simulation results with the numerical evaluation of~\eqref{eq:sinr_closed_form} presented in this subsection.
{We demonstrate that, for CFO models based on~\eqref{eq:Omega_approax}–\eqref{eq:exp_omega_sq_normal}, the conditional closed-form SINR in~\eqref{eq:sinr_closed_form} provides a reasonably accurate estimate of the measured SINR within small STD ranges.}
Conversely, for larger ranges of CFO STD, the closed-form approximate models in~\eqref{eq:Omega_approax}–\eqref{eq:exp_omega_sq_normal} become insufficiently accurate.
In such cases, accurately computing the SINR using~\eqref{eq:sinr_closed_form} requires evaluating the expectation terms $\mathbb{E}\{\Omega\}$ and $\mathbb{E}\{|\Omega|^2\}$ through Monte Carlo averaging over CFO realizations, rather than relying on the closed-form approximations.


\subsection{{Monte Carlo Simulation of SINR Performance}}
\label{subsec:sinr_closed-form_sim}

We simulate an OFDM transmission and reception chain based on {the 802.11 standard~\cite{802-11_standard} OFDM format.}
Out of \(N_{sc}=64\) subcarriers, we use \(48\) subcarriers for data transmission, \(4\) for pilot, and the rest as null subcarriers.
The model simulates \(M = 4\) dAPs with \(N = 16\) antennas per dAP, serving \(K = 4\) single-antenna users at the same time-frequency resource.
Users transmit their signals in uplink with the same power \(p_{\mrm{ul}}\), where the ratio of \(p_{\mrm{ul}}\) to the noise variance \(\sigma_z^2\) in the environment is \(20\) dB.
Users first send orthogonal uplink pilots.
These pilots are composed of two consecutive long training sequence (LTS) symbols based on 802.11 format and will be used for channel estimation and combining weight calculation.
The cyclic prefix length is \(L_{cp}=16\), and QPSK modulation will be used, which is more robust to CFO compared to higher-order modulations.
CFO is induced in the system in the form of realizations of the normal distribution as $\epsilon \sim \mathcal{N}(0,\beta^2)$.
After receiving signals and estimating channels, CBF is applied, and SINR and EVM are measured for each scenario.

\begin{figure}[!t]
\centering
\subfloat [{SINR vs $\beta$}] {
    \includegraphics[width=0.47\linewidth]{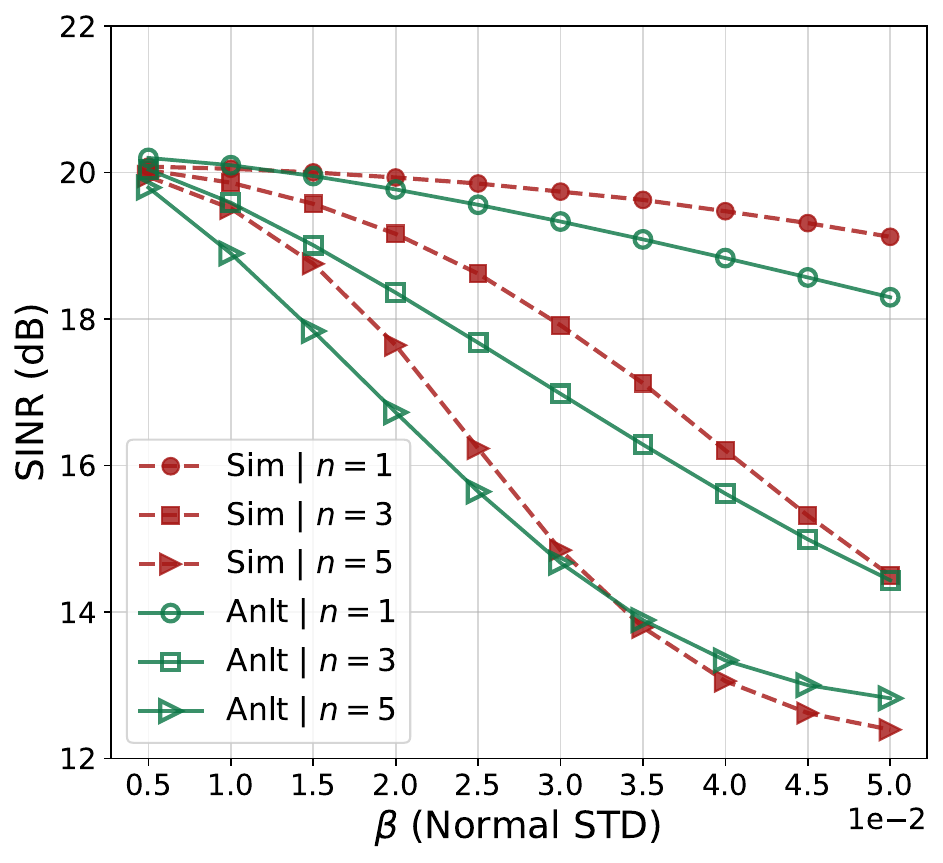}
    \label{fig:sim_sinr_std_normal}
    }
\subfloat [{EVM vs $\beta$}] {
    \includegraphics[width=0.47\linewidth]{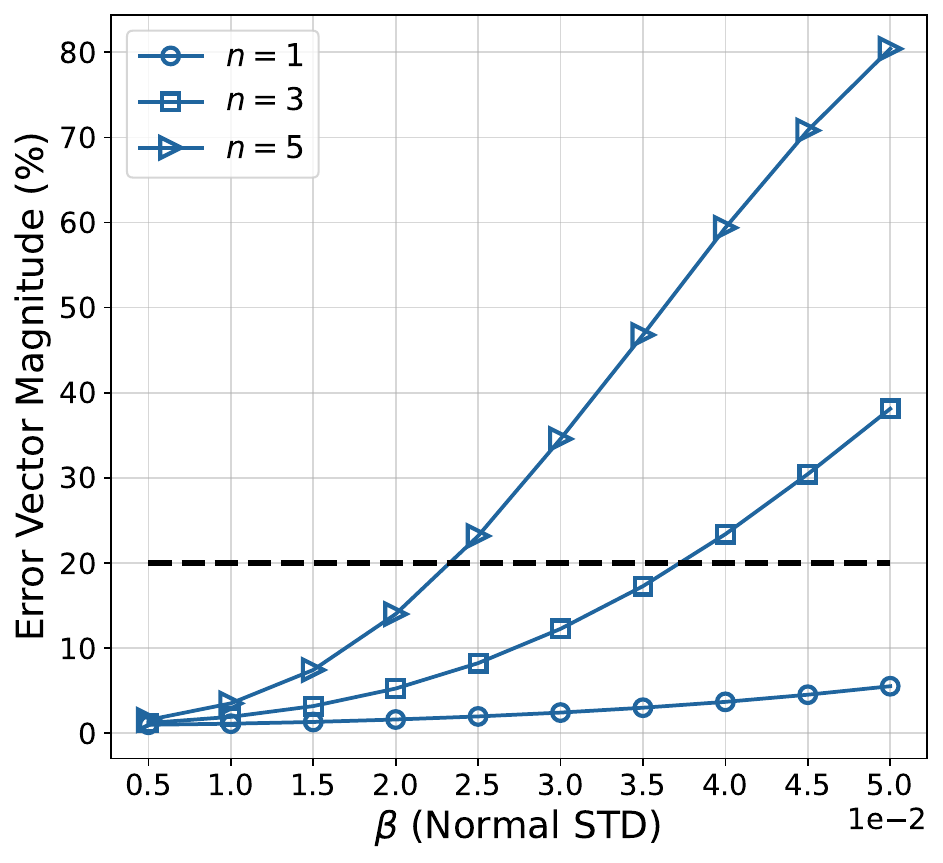}
    \label{fig:sim_evm_std_normal}
    }
\caption{{\textbf{(a)} Simulated SINR values (Sim) and analytical SINR estimates based on~\eqref{eq:sinr_closed_form} (Anlt) for conjugate beamforming (CBF). \textbf{(b)} Corresponding EVM results (in percentage) obtained from the same simulation setup.}}
\label{fig:sim_anlt_beta}
\end{figure}

{These simulations are conducted to compare the measured SINR, obtained by introducing CFO as random realizations, with the analytical SINR expression in~\eqref{eq:sinr_closed_form}, where the CFO effects are modeled using the closed-form approximations in~\eqref{eq:Omega_approax}–\eqref{eq:exp_omega_sq_normal}.}
A comparison is made in Fig.~\ref{fig:sim_sinr_std_normal}, where both the measured SINR based on the Monte Carlo simulation (labeled as Sim) and the analytical SINR based on \eqref{eq:sinr_closed_form} and \eqref{eq:Omega_approax}–\eqref{eq:exp_omega_sq_normal} (labeled as Anlt) are shown for CBF.
The CFO distribution is normal, with STD \(\beta\) ranging from \(0.5 \%\) to \(5 \%\).
For instance, when \(\beta = 2 \%\), according to the characteristics of the normal distribution, over \(99.7 \%\) of the normalized CFO (\(\epsilon\)) values are within \(3 \beta = 6 \%\) of the mean, which means \(\epsilon \in [-6\%, \, 6\%]\).
For a subcarrier spacing of \(\Delta f = 30\)~kHz, \(6\%\) normalized CFO corresponds to an offset of \(\delta_f = 0.06 \times \Delta f = 1.8\)~kHz from the true carrier frequency.
The ratio of this offset to the carrier frequency ($\delta_f / f_c$), measured in parts per billion (PPB), is equivalent to $750$~PPB, $600$~PPB, and $500$~PPB for the center frequencies of $2.4$~GHz, $3$~GHz, and $3.6$~GHz, respectively.
This range of CFO can significantly affect the performance of the system.
The results in Fig.~\ref{fig:sim_sinr_std_normal} illustrate that SINR drops with increasing \(\beta\), and this drop is faster for larger OFDM symbol indices.
This is because the phase rotation caused by CFO linearly scales with the symbol index \(n\) within an uplink slot with multiple OFDM symbols.
It also demonstrates that the SINR expression \eqref{eq:sinr_closed_form} with the CFO model \eqref{eq:Omega_approax}–\eqref{eq:exp_omega_sq_normal} gives an appropriate estimate of the average SINR for different \(\beta\) and \(n\) values.

Although we assess the SINR in our analytical framework, EVM is often used as a performance metric to evaluate empirical datasets~\cite{nakagawa_phase_noise, helfenstein_evm}.
SINR accounts for interference effects and serves to measure the channel quality indicator (CQI), which is utilized as the parameter for dynamically adjusting the modulation and coding scheme (MCS)~\cite{asilomar_mcs, bobrov_mcs}.
However, in a multi-user transmission scenario, measuring the power of IUI and separating it from the power of a desired signal and the power of noise is not feasible in most practical scenarios, subsequently making accurate SINR measurements difficult~\cite{brown_evm, marco_channel_dynamics}.
Due to the practical challenges of SINR measurement, EVM can be a viable alternative metric in real-world experiments because of the feasibility of measurement.
EVM can provide insight into the different sources of imperfections, including carrier leakage, IQ imbalance, non-linearity, and LO frequency errors~\cite{hisham_evm, liu_evm_est}.
Moreover, in several wireless communication standards, the requirements on EVM are defined as part of the standard, such as IEEE 802.11a~\cite{ieee_wlan} and IEEE 802.16e~\cite{ieee_wimax}, among others.

\begin{figure}[!t]
\centering
\subfloat [{SINR vs $n$}] {
    \includegraphics[width=0.47\linewidth]{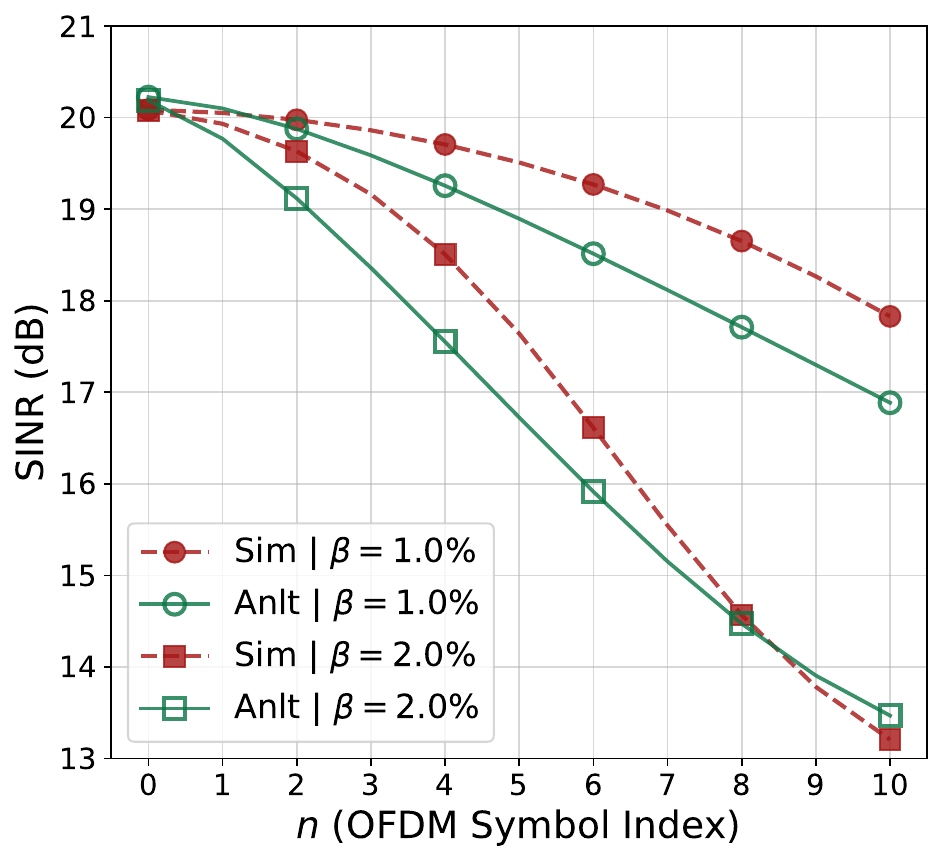}
    \label{fig:sim_sinr_n}
    }
\subfloat [{EVM vs $n$}] {
    \includegraphics[width=0.47\linewidth]{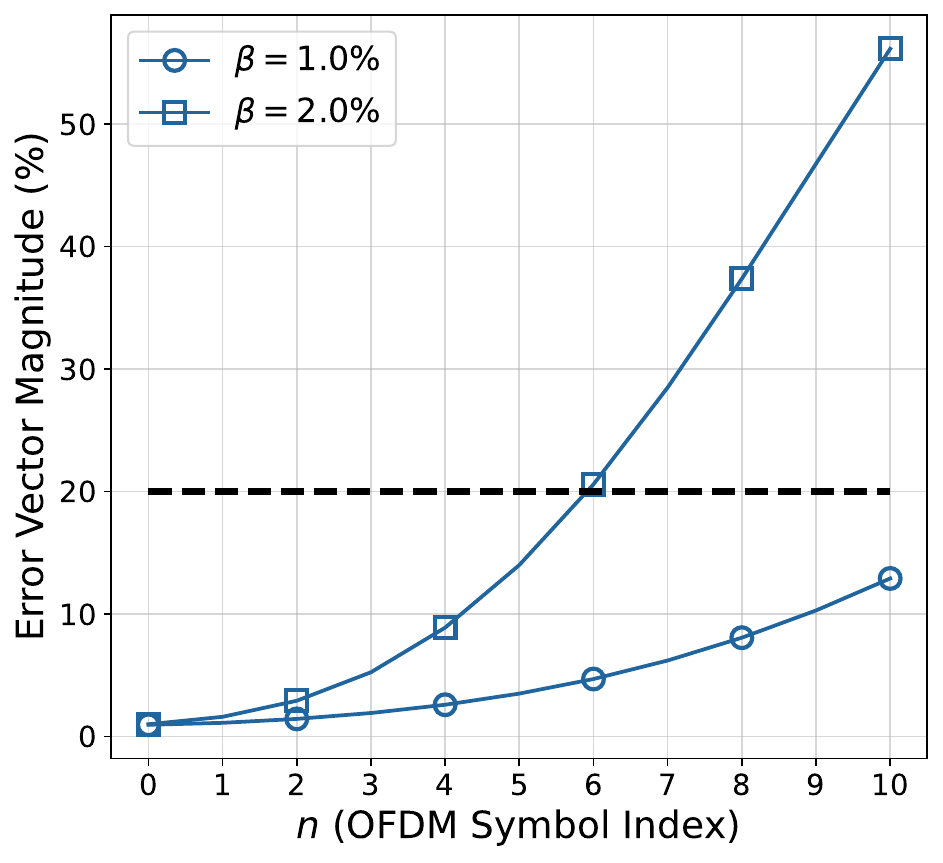}
    \label{fig:sim_evm_n}
    }
\caption{{\textbf{(a)} Simulated and analytical SINR and \textbf{(b)} simulated EVM results for $\beta = \{1\%, 2\%\}$ values, as functions of the OFDM symbol index $n$.}}
\label{fig:sim_anlt_n}
\end{figure}

{Accordingly, we evaluate the root mean square error vector magnitude ($\mathrm{EVM}_{rms}$) in our Monte Carlo simulations, as shown in Fig.~\ref{fig:sim_evm_std_normal}, where $\mathrm{EVM}_{rms}$ is defined as in~\cite{hisham_evm}:}
\begin{equation}
    \label{eq:evm_def}
    \textstyle \mathrm{EVM}_{rms} = \sqrt{\frac{1}{P_0 \, N_{\mrm{syms}}} \sum_{j=1}^{N_{\mrm{syms}}} \left|\mrm{s}_{tx}(j) - \mrm{s}_{rx}(j)\right|^2},
\end{equation}
where it is normalized to the average symbol energy of the constellation, \(P_0 = \frac{1}{\mrm{M}_{\mrm{mod}}} \sum_{i=1}^{\mrm{M}_{\mrm{mod}}} \left|\mrm{s}_i\right|^2\), to remove the dependency on the modulation order \(\mrm{M}_{\mrm{mod}}\). 
In \eqref{eq:evm_def}, \(\mrm{s}_{tx}(j)\) indicates the transmitted constellation symbol, \(\mrm{s}_{rx}(j)\) is the demodulated received symbol, {and \(N_{\mrm{syms}}\) indicates the total number of transmitted symbols.}
Using this definition, we derived EVM results presented in Fig.~\ref{fig:sim_evm_std_normal}, for the simulation scenarios in Fig.~\ref{fig:sim_sinr_std_normal}.
It is demonstrated in Fig.~\ref{fig:sim_evm_std_normal} that EVM increases rapidly by increasing the STD of the CFO distribution.
As illustrated, performance degradation is higher for larger symbol indices.
We can infer from this result that if a maximum EVM value is considered for a system (for example \(20 \%\)), then the maximum tolerable CFO and the size of the uplink slot (in terms of the number of OFDM symbols) are two dependent parameters to be designed accordingly for the system.
Based on results in Fig.~\ref{fig:sim_evm_std_normal}, if we want to have \(\mrm{EVM} \leq 20\%\) and the uplink slot contains \(n=5\) OFDM symbols, then we need to have \(\beta < 2.5\%\), meaning the normalized CFO must almost always remain in the range of \([-7.5\%, \, 7.5\%]\) of the subcarrier spacing.
Accordingly, for \(\epsilon = 7.5\%\), a subcarrier spacing of \(\Delta f = 30\) kHz allows us to tolerate CFO values up to \(\delta_f = 2.25\) kHz, while \(\Delta f = 60\) kHz makes the system resilient to CFO up to \(\delta_f = 4.5\) kHz.
Our analytical framework and SINR expressions can then be applied in this setting to quantify the SINR and analyze its dependence on the statistical properties of CFO.

We can further apply this analysis for a system that has a maximum tolerable CFO.
Assuming that the maximum CFO is \(\delta_f = 1.8\) kHz, it amounts to \(\epsilon = 6\%\) with a subcarrier spacing of \(\Delta f = 30\) kHz and \(\epsilon = 3\%\) with \(\Delta f = 60\) kHz.
Considering a normal distribution for CFO, based on the STD rule, the maximum STD for these two cases is \(\beta = 2\%\) and \(\beta = 1\%\), respectively.
We provided both SINR and EVM results for these two values of \(\beta\)
versus OFDM symbol index (\(n\)) values in Fig.~\ref{fig:sim_anlt_n}.
In this figure, $n = 0$ corresponds to the time instant immediately after synchronization and CFO compensation, where no residual phase offset is present.
After that, phase rotation imposed by CFO scales over the OFDM symbols within the uplink slot.
Fig.~\ref{fig:sim_sinr_n} illustrates a decline in SINR from \(20\) dB to approximately \(17\) dB over a span of \(10\) OFDM symbols with \(\beta = 1\%\) and to \(13\) dB with \(\beta = 2\%\).
Furthermore, Fig.~\ref{fig:sim_evm_n} demonstrates that when \(\beta = 2\%\), the EVM exceeds \(20\%\) after the \(6\)-th OFDM symbol and exhibits near-linear growth with the symbol index $n$.

This analysis showcases that our SINR-CFO framework can be utilized to assess and define system parameters such as the maximum tolerable CFO, desired EVM, transmission slot size, and subcarrier spacing according to specific application demands and system requirements.
Moreover, the simulation results offer important technical insights: as the CFO variance increases, inter-dAP phase misalignment accumulates over time, leading to degraded signal combining and reduced SINR.
This degradation is particularly pronounced for later OFDM symbols (larger $n$), where the accumulated phase rotation induced by CFO becomes more severe.
The close alignment between analytical and simulated SINR for small CFO variances validates the effectiveness of our proposed model in practical regimes, while deviations at higher variances highlight the limitations of the closed-form approximation in such regimes, with high mobility or severe hardware impairments.

\begin{figure}[t]
    \centering
    \includegraphics[width=2.8in]{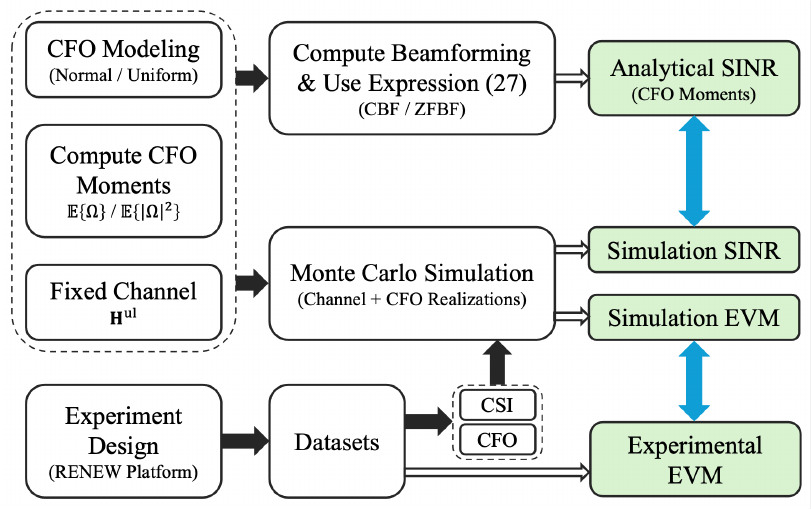}
    \caption{{Block diagram of the proposed framework for evaluating the impact of CFO on beamforming performance. Blue arrows indicate comparison points.}}
    \label{fig:block_diagram}
\end{figure}

To provide a high-level overview of our approach, Fig.~\ref{fig:block_diagram} illustrates the key components of the proposed framework and their interactions.
The framework includes the analytical derivation of SINR as a function of CFO statistics, Monte Carlo simulation with varying channel and CFO realizations, and real-world experiments using the RENEW testbed.
The analytical part takes as input the statistical properties of residual CFO and a fixed uplink channel realization to compute the SINR expression accounting for CFO-induced degradation.
The same inputs are used in Monte Carlo simulations to validate the analytical model by computing the SINR values.
In addition, the experimental datasets are leveraged to validate both the simulation and analytical results.


\section{Experiments}
\label{sec:experiment}

In this section, we first introduce the RENEW~\cite{renew} indoor massive MIMO platform.
We then describe the experimental setup and outline the datasets collected under various experimental conditions.
Furthermore, we showcase several experimental results derived from these datasets along with comparisons with the analytical framework discussed in~\S~\ref{sec:analytical}.


\subsection{Hardware Setup}
\label{subsec:hw_setup}

We use the RENEW programmable massive MIMO platform~\cite{renew} to perform experiments.
The RENEW base station uses the Iris SDR~\cite{shepard2020argosnet} as its building block, with each SDR supporting two RF chains.
The base station includes a central hub that by default distributes clock and time trigger signals to up to 48 Iris SDRs (96 antennas) for time and frequency synchronization.
At each Iris, both the RF and sampling clocks are derived from the clock distributed by the hub.
Therefore, all antennas are locked at the carrier frequency $f_c$ in the form of a perfectly synchronized array.
In this scenario, perfect time and frequency synchronization among antennas is achieved via the central hub.
We denote this operational mode of the SDRs as \textit{HUB-mode}. 
An Iris can also be used in a standalone mode to emulate a user.
The RENEW platform is designed to operate at the CBRS spectrum ranging from \(3.55\)-\(3.7\) GHz.

Apart from receiving an external clock, each Iris is equipped with its own LO, a super-TCXO with a nominal $\pm 100$ PPB accuracy for the sub-\(6\) GHz band. 
Through firmware settings, the Iris can be configured to use its own LO to generate the RF clock or to use the external clock from the central hub.
To emulate a distributed antenna array with independent LOs, each Iris in the base station can be programmed to use its own LO, introducing a random CFO between the antennas.
In this operation mode, the antennas still receive the time trigger signal from the central hub as their time reference.
{Time synchronization in distributed systems can be achieved through well-established protocols, such as IEEE 1588 Precision Time Protocol (PTP) or GPS-based methods~\cite{sami2010time_sync, schwartz2022time_sync, chen2019time_sync, zhang2023time_sync}.}
Distributing the time trigger across radios ensures time alignment, allowing us to isolate and focus solely on the impact of CFO on system performance.
We denote the operation mode in which SDRs use their LO to derive the RF clock as \textit{LO-mode}.

\begin{figure}[!t]
\centering
\subfloat [dAPs] {
    \includegraphics[width=0.45\textwidth]{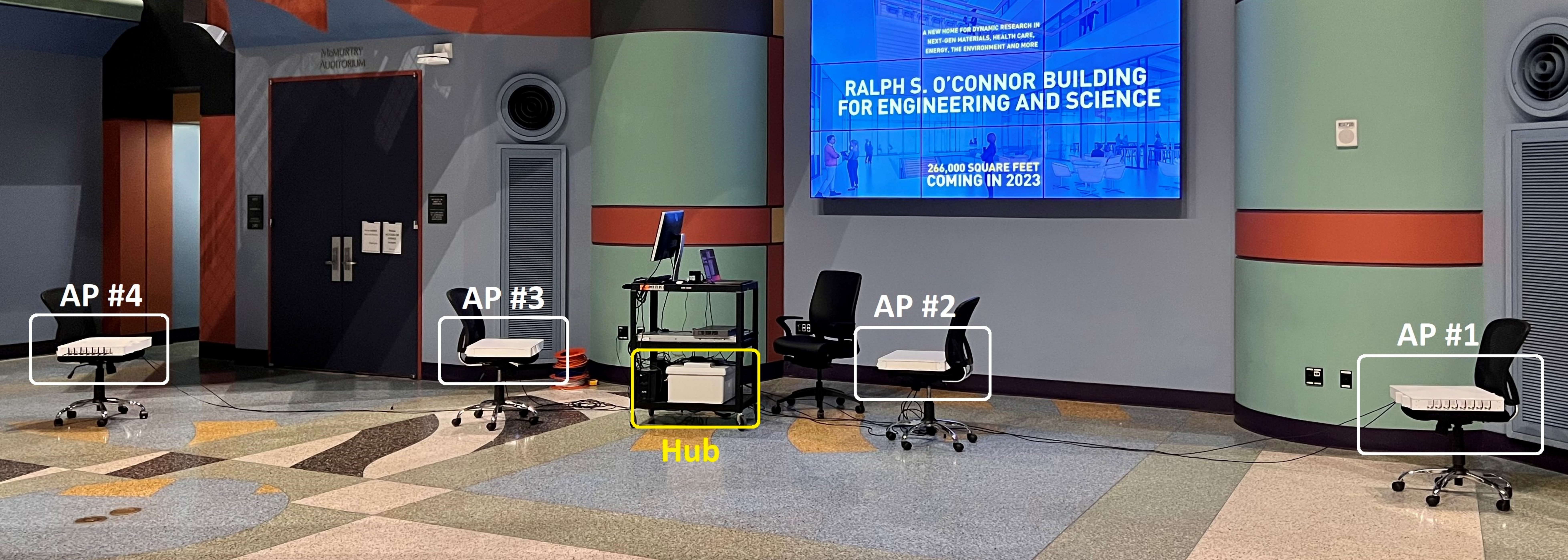}
    \label{fig:exp_setup_ap}
    }
    
\subfloat [UEs] {
    \includegraphics[width=0.45\textwidth]{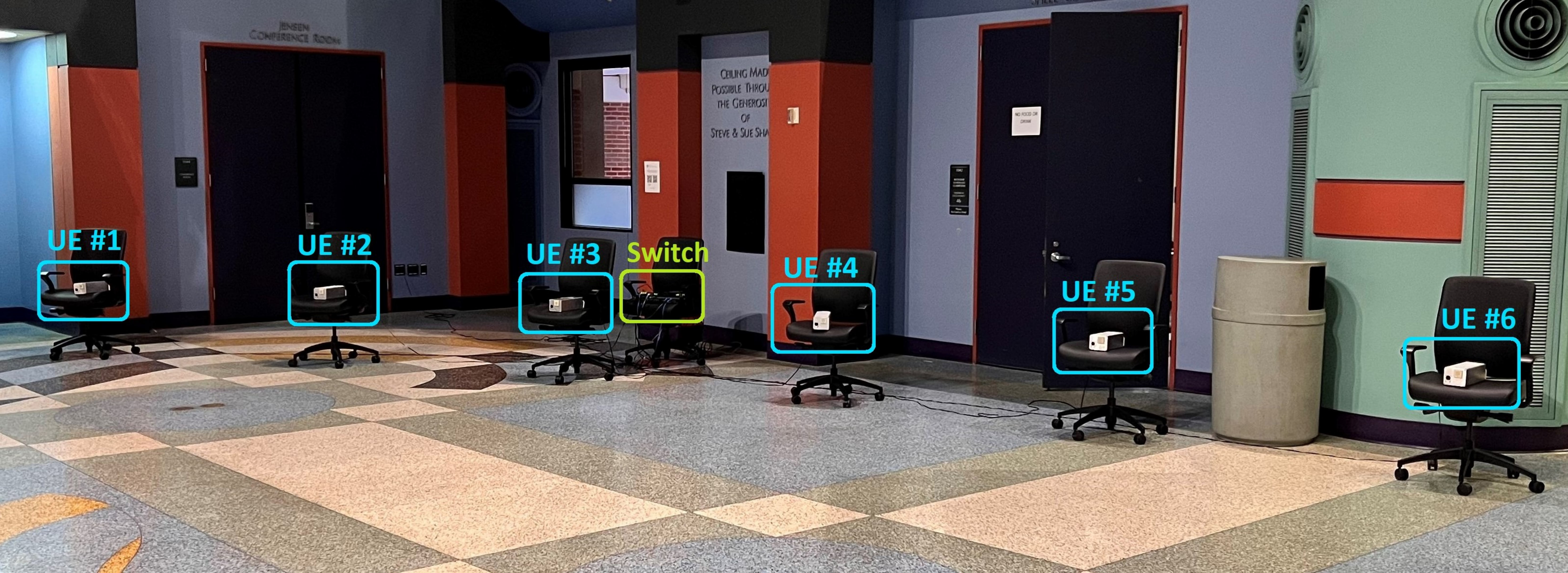}
    \label{fig:exp_setup_ue}
    }
\caption{{\textbf{(a)} Placement of the RENEW BS chains as dAPs in the distributed configuration and \textbf{(b)} placement of standalone Iris SDRs representing users.}}
\label{fig:exp_setup}
\end{figure}


\subsection{Experiment Design}
\label{subsec:exp_design}

For our experimental setup, we utilize an indoor RENEW massive MIMO base station, equipped with four chains of Iris SDRs, consisting of 64 antennas in total.
Each chain consists of 8 interconnected SDRs with dual-polarized antennas, forming a linear array that supports a total of 16 streams.
All of these chains are connected to the central hub via a wired fiber link.
These chains are considered as APs and are employed in two physical configurations: co-located and distributed.
The clock signal in each AP is achieved through a daisy chain approach, where the initial SDR (head) of the chain disseminates the clock to all other SDRs in that AP.
The clock at the head SDR can either be received from the hub, referred to as HUB-mode, or derived from its own independent LO, referred to as LO-mode.
The experiments were conducted in an indoor hall with a radius of approximately 17 meters. We collected datasets across four scenarios, outlined below:
\begin{enumerate}
    \item \textbf{Cent-HUB} (centralized \& HUB-mode):
    \newline All four APs are stacked on top of each other, forming a 2D co-located 64-antenna massive MIMO array. APs are also perfectly synchronized in time and frequency through the fiber connection to the hub. 
    \item \textbf{Cent-LO} (centralized \& LO-mode):
    \newline APs are placed in the centralized form while the head of each AP is toggled to derive the carrier and sampling frequencies from its own LO and distribute it among other antennas of the AP. Therefore, we will have four 16-antenna APs with four independent LOs that do not perform any extra frequency synchronization step. The radios still receive the time trigger signal from the hub. 
    \item \textbf{Dist-HUB} (distributed \& HUB-mode):
    \newline APs are located 3 meters away from each other (called dAPs in this case) to emulate a physically distributed massive MIMO array, and are synchronized via the hub.
    \item \textbf{Dist-LO} (distributed \& LO-mode):
    \newline APs are distributed (dAPs) and operating in LO-mode.
\end{enumerate}

Within each scenario, we use six standalone Iris SDRs as users, evenly spaced to maintain equal power levels, as illustrated in Fig.~\ref{fig:exp_setup}.
All users are in the line-of-sight (LOS) of the AP antennas.
We collected channel and uplink data in configurations that include \(64\times 1\), \(64\times 2\), \(64\times 4\), and \(64\times 6\) antennas.
For each configuration, multiple subsets of users are selected to conduct the experiments.
Both the uplink pilot and data samples for all of the above-mentioned scenarios in all possible configurations are included in our datasets.
One can leverage the datasets to particularly explore the impact of the distance between UEs, between dAPs, channel correlations, and residual CFO from independent LOs on multi-user beamforming performance.
In this paper, we emphasize the performance of D-MUBF by concentrating on the fourth scenario, Dist-LO, which involves geographically distributed APs operating with their individual LOs.
Fig.~\ref{fig:exp_setup} illustrates the placement of dAPs and UEs in this scenario. 

\begin{table}[!t]
\caption{Detailed Experiment Parameters}
\label{tab:exp_params}
\centering
\begin{tabular}{|c|c|}
\hline
    \textbf{Parameter} & \textbf{Value} \\
    \hline
    Total antennas (\(M\times N\))  &   64   \\
    Antennas per AP (\(N\))         &   16   \\
    Number of users (\(K\))         &   1, 2, 4, 6 \\
    Environment                     &   Indoor \\
    AP-user distance                &   14 m \\
    Height of each AP               &   0.6 m (2 ft) \\
    Height of each user             &   0.6 m (2 ft) \\
    Distance b/w dAPs               &   0 m / 3 m \\
    Distance b/w users              &   2 m \\
    Temperature                     &   65 F \\
    
    \hline

    Operating frequency         &   3.6 GHz \\
    Bandwidth                   &   5 MHz / 10 MHz \\
    {Antenna gain}                &   {5 dBi} \\
    Modulation                  &   QPSK \\
    Training pilot              &   802.11 LTS \\
    \# of Training sequences    &   2 \\
    \# of OFDM subcarriers      &   64 (48 data sc) \\
    CP Length                   &   16 \\
    OFDM data symbols per slot  &   10 \\
    \hline
\end{tabular}
\end{table}

We use the RENEWLab real-time software~\cite{RENEWLab} to collect datasets in all of the configurations.
The RENEWLab software runs on a host server that is connected to the RENEW BS through fiber. 
Through a configurable frame schedule, RENEWLab software orchestrates the transmission and reception of pilot and data OFDM symbols at users and from the BS antennas, respectively.
Within each frame, one of the BS antennas transmits a synchronization beacon so that users can synchronize their frame time boundaries with the BS frame. 
We set up the frame schedule for the users so that each user sends time-orthogonal pilots and then all simultaneously transmit data OFDM symbols.
We use the 802.11 LTS signal as the pilot symbol, which is used for channel estimation and beamforming matrix calculation.
The format of the data OFDM symbols also follows the 802.11 {standard~\cite{802-11_standard}}, where the FFT size is $64$. Out of the $64$ subcarriers, $48$ subcarriers carry data, $4$ are pilot subcarriers, and the remaining are null subcarriers.
The pilot subcarriers are used for residual phase offset correction.
We try QPSK for modulating data subcarriers.
The detailed parameters of the experiments are shown in Table \ref{tab:exp_params}.
The RENEWLab framework enables the recording of pilot and data OFDM symbols received at BS antennas in a dataset file with HDF5 format~\cite{hdf5}.
The HDF5 datasets can be processed offline with the Python-based post-processing library in the RENEWLab software.
Our data collection effort resulted in over 250 HDF5 files, with each file comprising 2000 frames, and each frame containing pilot symbols for each user followed by $10$ data OFDM symbols.
Datasets are available on~\cite{dist_mu_mimo_datasets} for further research purposes.


\subsection{Results for Experimental Validation}
\label{subsec:exp_results}

We begin by processing the uplink pilot samples received at the dAPs to estimate the user-to-dAP channels.
Additionally, pilot samples from individual users in each experimental scenario are used to estimate the CFO between each user and dAP.
To compute the inter-dAP CFO, we measure the CFO values at all receive antennas relative to a designated reference antenna.
The empirically estimated channel and CFO values extracted from the datasets are then incorporated into both our Monte Carlo simulation and analytical frameworks for performance comparison.

\subsubsection{Hardware CFO Analysis}
\label{subsubsec:hw_cfo}

\begin{figure}[!t]
\centering
\subfloat [user-dAP CFO Histogram] {
    \includegraphics[width=0.23\textwidth]{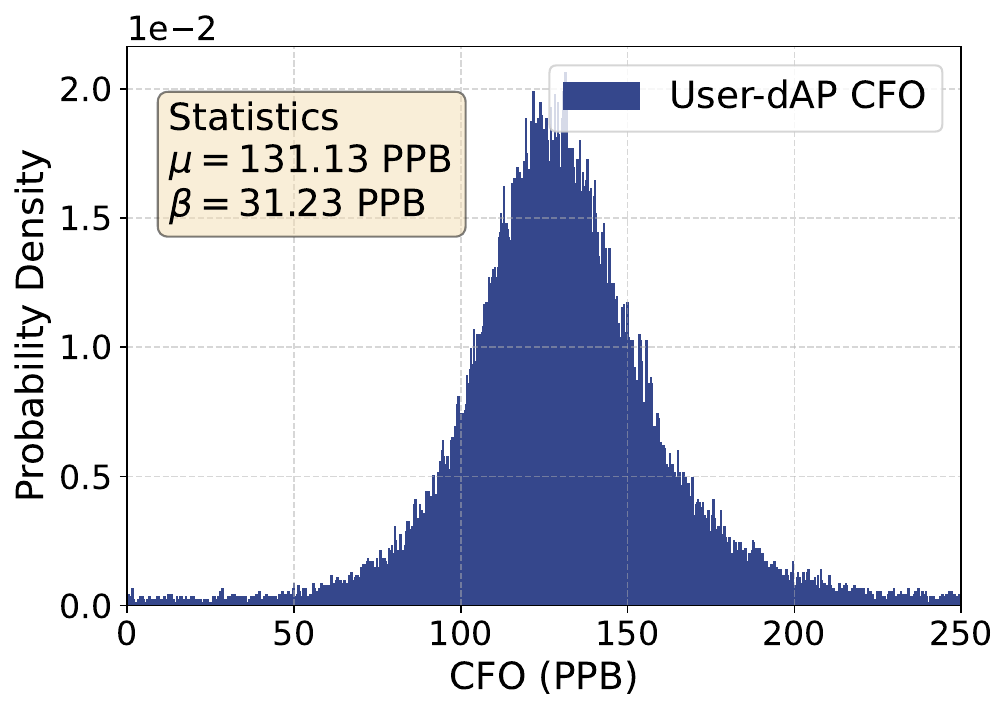}
    \label{fig:su_CFO_hist}
    }
\subfloat [user-dAP CFO Mean Values] {
    \includegraphics[width=0.23\textwidth]{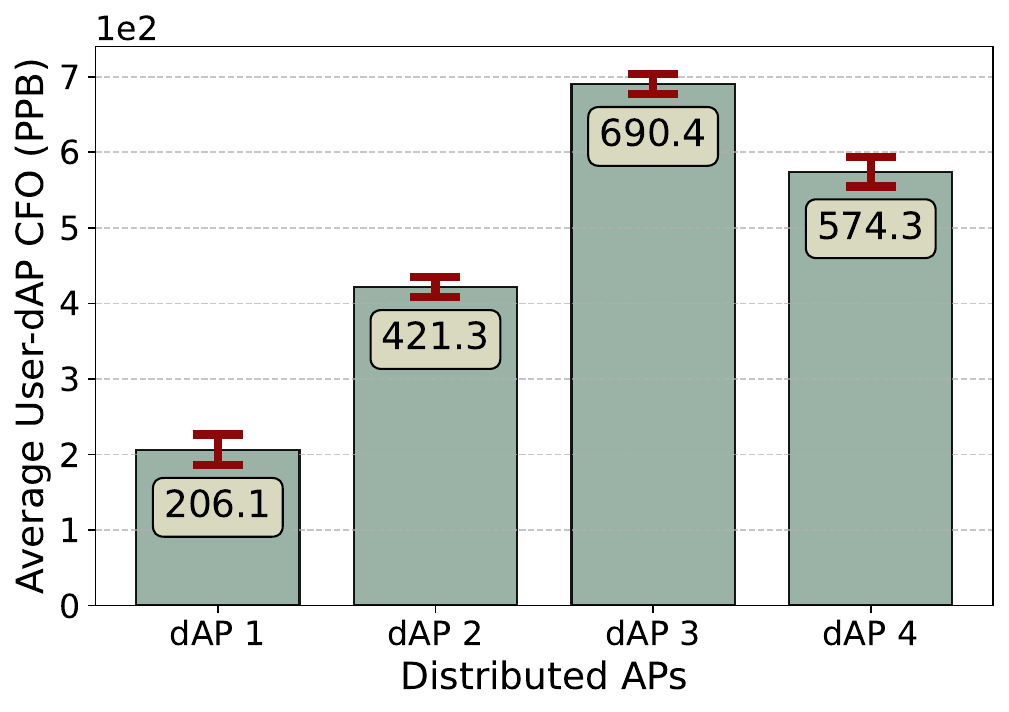}
    \label{fig:cfoVsAnts_64x1}
    }
\caption{{\textbf{(a)} Distribution of the user-dAP CFO estimated using pilots in the HUB-mode when there is no inter-dAP CFO and \textbf{(b)} average values of the estimated CFO between users and dAP antennas in the LO-mode, for \(64 \times 1\) configuration. The CFO is measured in PPB relative to the operating frequency of \(3.6\) GHz, highlighting distinguishable frequency offsets among the dAPs.}}
\label{fig:exp_cfo}
\end{figure}

Due to manufacturing variations, oscillators used in Iris SDRs have random offsets from the true operating frequency.
To obtain an estimate of the CFO between the transmitter and each receiving antenna, we process the pilot samples received by the dAP antennas from each user.
In the HUB-mode, where the central hub provides timing and frequency synchronization for all dAP antennas, we anticipate perfect synchronization, resulting in zero inter-dAP CFO.
Nevertheless, in this configuration, each user derives the carrier frequency using its independent LO, introducing a random CFO between the user and the dAPs.
To quantify this CFO, we analyzed pilot samples collected in the Dist-Hub scenario.
The histogram illustrating the estimated CFO values between one of the users and the dAPs is presented in Fig. \ref{fig:su_CFO_hist}, with the CFO expressed in PPB relative to the operational frequency, set at \(3.6\) GHz in our experiments.
The histogram in Fig.~\ref{fig:su_CFO_hist} demonstrates that, in the HUB-mode, the mean user-to-dAP CFO is approximately \(131\) PPB, equivalent to \(470\) Hz.
Considering that the subcarrier spacing in this experiment is \(\Delta f = 78.125\) kHz, the normalized CFO becomes \(\epsilon = 0.6\, \%\).

In the Dist-LO scenario, where each dAP employs an independent LO to derive carrier frequency, we anticipate the presence of inter-dAP CFOs.
In this scenario, we measure the average estimated CFO values with one user transmitting and all dAP antennas receiving.
Fig.~\ref{fig:cfoVsAnts_64x1} demonstrates the mean user-dAP CFO values across all four dAPs.
Within each dAP (comprising 16 antennas), the CFO values are relatively consistent across antennas, indicating effective intra-dAP frequency synchronization. However, the average CFO values differ across dAPs, with the difference ranging from approximately $100$ PPB to $300$ PPB, highlighting the presence of inter-dAP CFO.
Throughout this experiment, the largest CFO was observed between the reference user and dAP 3, as shown in Fig.~\ref{fig:cfoVsAnts_64x1}, with a magnitude of $700$ PPB. This corresponds to a normalized CFO of approximately $\epsilon = 3.23\%$.

To enable a meaningful comparison between analytical and experimental results, {we use our Monte Carlo simulation model to replicate the transmission process under the same parameter settings as those used in the experiments.
We extract estimated channels and inter-dAP CFO statistics from the experimental data and incorporate them into our simulation model.
It is assumed that users are locked to the desired carrier frequency, and the only residual CFO present is the inter-dAP CFO.
This residual CFO is modeled as a normal random variable with zero mean and $\beta^2$ variance.
For each experimental scenario, the parameter $\beta$ is empirically estimated by analyzing the corresponding dataset.}
Next, we present and compare analytical and experimental results on beamforming performance across multiple experimental scenarios.

\subsubsection{Beamforming Performance Analysis}
\label{subsubsec:bf_performance}

In the single-user experiment, we have one single-antenna user transmitting in the uplink and four 16-antenna dAPs receiving the signal.
The datasets collected using each individual user in a \(64 \times 1\) configuration are analyzed for producing the results.
We use EVM-SNR in dB as our performance metric to evaluate the experimental datasets, which is defined as
$   \text{EVM-SNR} = 10 \, \log_{10} \left(1/\text{EVM}_{rms}\right),
$ 
for $\text{EVM}_{rms}$ defined in \eqref{eq:evm_def}.

\begin{figure}[!t]
    \centering
    \includegraphics[width=2.85in]{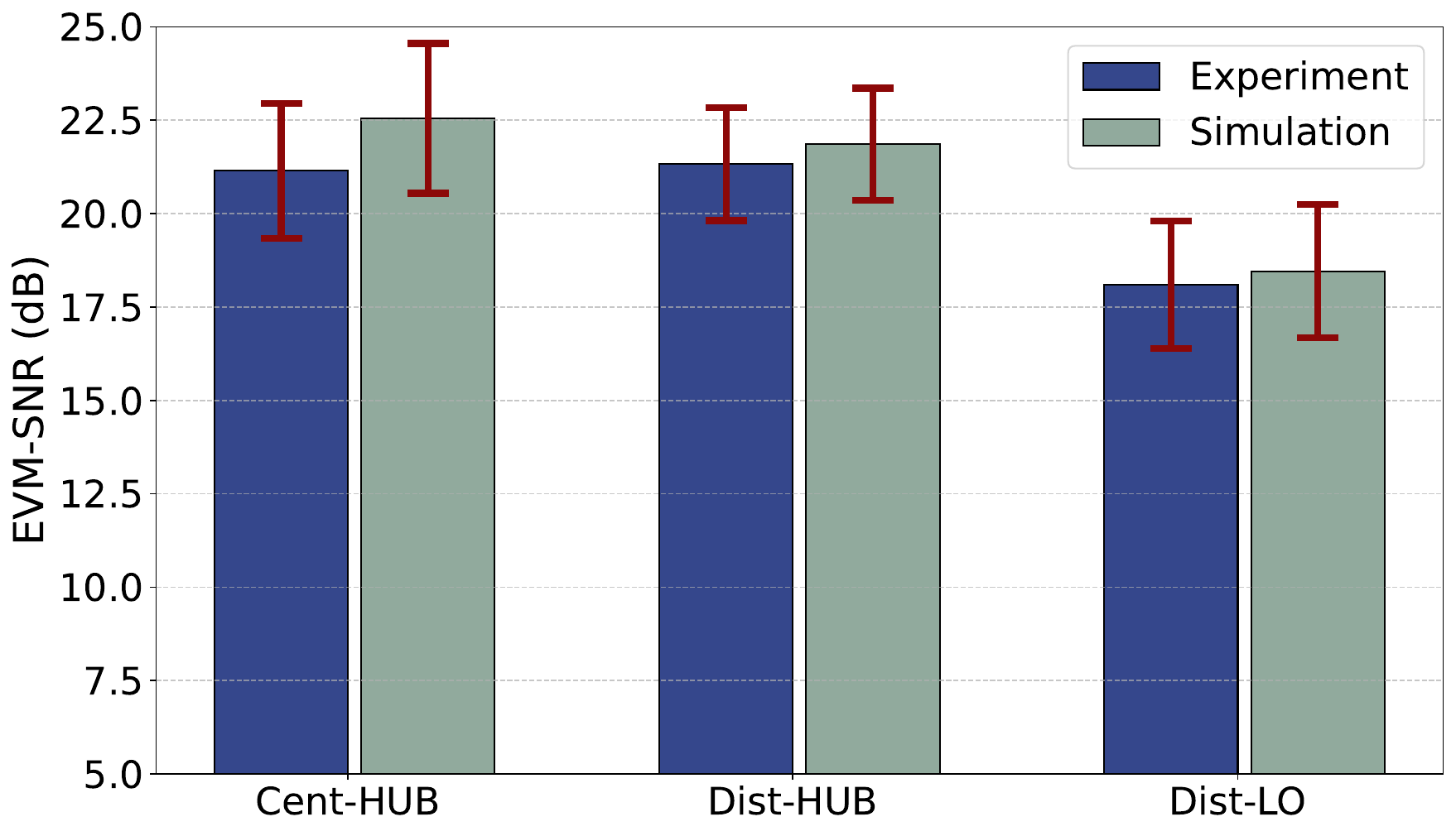}
    \caption{{Experimental and simulation EVM-SNR for \(64\times 1\) configuration in different experiment scenarios. HUB-mode refers to perfect synchronization via the hub, and LO-mode refers to using independent LOs at each AP.}}
    \label{fig:evm_exp_sim_su}
\end{figure}

We analyze the data samples within the datasets to derive the average EVM-SNR for the single-user configuration across various scenarios.
We apply the ZFBF using the received data across all antennas.
For single-user receive beamforming, ZFBF is functionally equivalent to CBF.
Fig.~\ref{fig:evm_exp_sim_su} presents the corresponding results, including both Monte Carlo simulations (using matched parameter settings) and experimental measurements of the average EVM-SNR.
For each experimental scenario, we replicate the setup in simulation by incorporating the estimated channels and CFO statistics. Using this Monte Carlo simulation model, we obtain both EVM-SNR results for comparison with experimental measurements and SINR values for comparison with the analytical framework.

The presentation of simulation and experimental results in Fig.~\ref{fig:evm_exp_sim_su} enables a direct comparison across different scenarios.
As shown in the figure, the average EVM-SNR values for both centralized and distributed setups operating in HUB-mode are comparable.
In contrast, the Dist-LO scenario exhibits a noticeable performance degradation of approximately $4$ dB relative to the Dist-HUB configuration.
Moreover, a compelling observation is that the proposed simulation model, which models the inter-dAP CFO as a zero-mean normal distribution, provides reliable and consistent estimates of the average EVM-SNR across all evaluated scenarios.

Across various multi-user configurations, where different user subsets were selected to capture the effects of high and low spatial correlation in beamforming, we measured the average estimated CFO between the dAPs.
The results show that inter-dAP CFO values in these multi-user experiments ranged from $150$ to $350$ PPB, while intra-dAP synchronization remained consistently maintained, as expected.
Notably, the maximum observed CFO across all multi-user measurements approached approximately \(900\) PPB, which corresponds to a normalized CFO of \(\epsilon = 4.15\, \%\) when considering a subcarrier spacing of \(\Delta f = 78.125\) kHz.
These results are consistent with our assumption of negligible ICI impact, given that normalized inter-dAP CFO variations remain within the \(5\%\) range.

\begin{figure}[!t]
    \centering
    \includegraphics[width=2.85in]{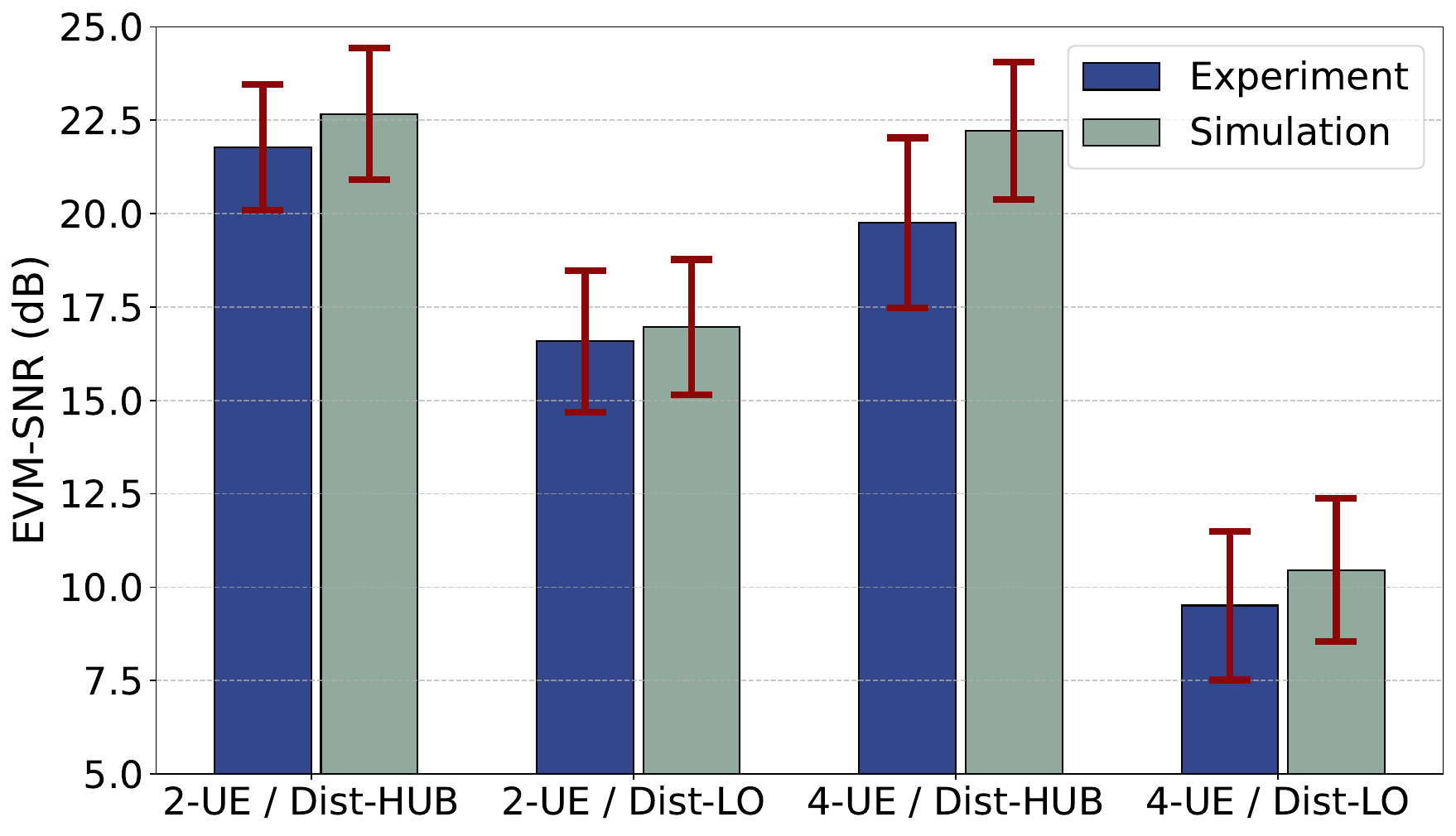}
    \caption{{Comparing experimental and simulation EVM-SNR results for \(64\times 2\) and \(64\times 4\) configurations in both the HUB-mode and LO-mode scenarios.}}
    \label{fig:evm_exp_sim_mu}
\end{figure}

For two multi-user configurations of \(64 \times 2\) and \(64 \times 4\) with distributed placement of APs, we processed the received data samples in both the HUB-mode and LO-mode, employing ZFBF to mitigate inter-user interference and recover the uplink data from each user.
In the corresponding Monte Carlo simulations, we used the estimated CSI and CFO values obtained from the experimental measurements.
The average per-user EVM-SNR measurements for these multi-user scenarios are presented in Fig.~\ref{fig:evm_exp_sim_mu}, comparing the performance of Dist-HUB versus Dist-LO scenarios.
As illustrated in Fig.~\ref{fig:evm_exp_sim_mu}, the performance drop between the Dist-HUB and Dist-LO scenarios is notable.
{In the two-user configuration, there is a reduction of approximately \(5\) dB in the average EVM-SNR, while this degradation escalates to approximately \(9\) dB in the four-user setup.
This performance gap highlights the increased sensitivity of higher-order multi-user MIMO configurations to synchronization accuracy.
The greater vulnerability arises from the shared utilization of time and frequency resources by multiple users, making the received signal susceptible to the aggregate effects of CFOs from different user-dAP pairs.
As a result, conventional CFO correction techniques, which typically assume a single dominant offset, become ineffective in distributed multi-user transmissions, where CFO contributions accumulate across spatially separated links.}

Comparing different MIMO configurations of $64 \times 1$, $64 \times 2$, and $64 \times 4$, the average estimated inter-dAP CFO values in all cases fall within the range of $150$ PPB to $350$ PPB, corresponding to normalized CFO values of $\epsilon = 0.69\%$ and $\epsilon = 1.61\%$, respectively, for $\Delta f = 78.125$ kHz. Based on the results in Fig.~\ref{fig:evm_exp_sim_su}, this CFO range leads to a performance degradation of approximately $4$ dB in EVM-SNR for the single-user scenario. In the two- and four-user configurations, as demonstrated in Fig.~\ref{fig:evm_exp_sim_mu}, the degradation increases to around $5$ dB and $9$ dB, respectively, when using ZFBF.

\subsubsection{Connection to the Analytical Framework}
\label{subsubsec:exp_analytical_comparison}

\begin{figure}[!t]
    \centering
    \includegraphics[width=2.95in]{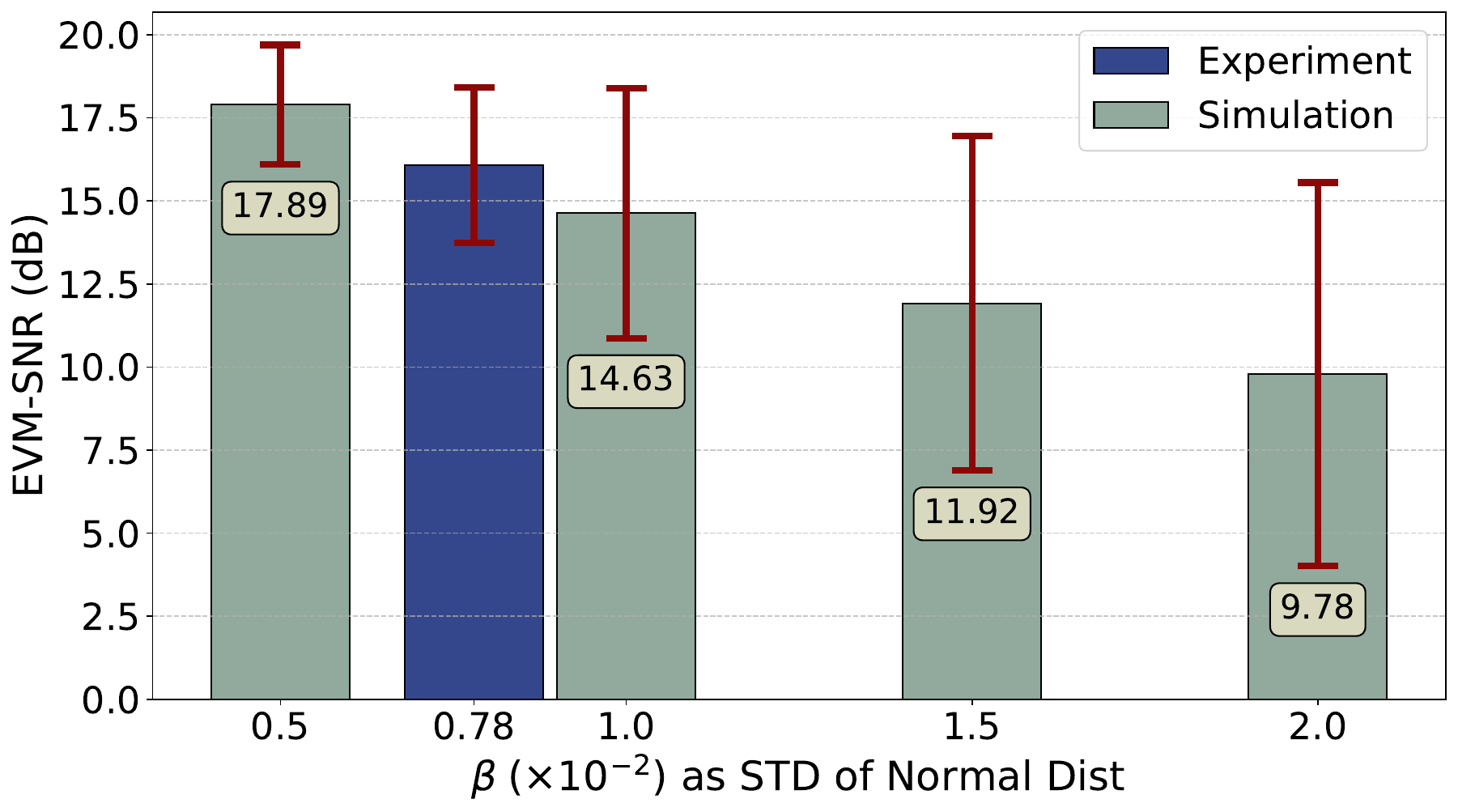}
    \caption{{Comparing experimental EVM-SNR for the \(64\times 2\) configuration to simulation results with multiple values of CFO STD. The empirical STD \(\beta = 0.78\times 10^{-2}\) ($0.78$\%) was estimated from the corresponding experiment.}}
    \label{fig:evm-snr_beta_sim}
\end{figure}

Based on the results in Fig.~\ref{fig:evm_exp_sim_su} and Fig.~\ref{fig:evm_exp_sim_mu} with ZFBF, we observed that our simulation model is able to effectively reproduce the EVM-SNR values by modeling the inter-dAP CFO as a zero-mean normal random variable, with the STD estimated from the experiments.
To validate this model for CBF and assess the behavior of the EVM-SNR for different values of the CFO STD \(\beta\), we perform a Monte Carlo simulation for the \(64\times 2\) configuration in the Dist-LO scenario.
We use the estimated channels from the corresponding experiment and use the normal distribution of CFO with four different STD values.
Using CBF, we evaluate the average EVM-SNR under each setting.
The results are presented in Fig.~\ref{fig:evm-snr_beta_sim}, where the green bars represent simulation results for varying $\beta$ values, and the blue bar shows the experimental result for the corresponding $64 \times 2$ configuration with the estimated $\beta$ value.

For the experimental result in Fig.~\ref{fig:evm-snr_beta_sim}, using pilot samples from the corresponding \(64\times 2\) experiment, we estimate the STD of the inter-dAP CFO, which resulted in the value of \(\beta = 0.78 \times 10^{-2}\). 
As demonstrated in Fig.~\ref{fig:evm-snr_beta_sim}, our simulation model yields average EVM-SNR values ranging from $17.9$ dB to $14.6$ dB as $\beta$ increases from $0.5 \times 10^{-2}$ to $1 \times 10^{-2}$, respectively.
Our experimental result shows the average EVM-SNR of \(16\) dB with the estimated \(\beta\) in this range.
Thus, by comparing the experimental and simulation results, we can confirm the effectiveness of our simulation model in replicating the experiments, as well as validating the inter-dAP CFO model and the estimated inter-dAP CFO statistics.
{Furthermore, the validated simulation model is used to support our analytical framework and the SINR expression in~\eqref{eq:sinr_closed_form}. Specifically, Figs.~\ref{fig:sim_anlt_beta} and~\ref{fig:sim_anlt_n} demonstrate close agreement between the analytical and simulated SINR results across varying CFO standard deviations, thereby confirming the accuracy and effectiveness of the proposed framework.}

{In summary, the experiments were conducted on the RENEW indoor massive MIMO testbed~\cite{renew}, using 64 SDRs grouped into four dAPs with 16 antennas each. The system operated in the CBRS band using OFDM with 64 subcarriers and $\Delta f = 78.125$ kHz of spacing. We evaluated both single-user and multi-user uplink configurations under HUB-mode and LO-mode synchronization, with users remaining stationary during each experiment. For every scenario, 2000 transmission frames were recorded, and channel estimates and inter-dAP CFO values were extracted from the received pilot signals. The datasets are available at~\cite{dist_mu_mimo_datasets}, and the corresponding codebase is available at\cite{RENEWLab}.
These empirical values were incorporated into our simulation and analytical models for comparison. The results confirm that our CFO-aware SINR framework and simulation model provide accurate predictions of beamforming performance, particularly
in capturing the degradation caused by inter-dAP CFO in distributed multi-user settings.}


\subsection{Discussion}
\label{subsec:discussion}

The insights gained from analyzing the CFO profile, including its distribution and variance, can guide the design of key system parameters.
These parameters might include subcarrier spacing, the number of beamforming antennas, and the selected beamforming approach, all of which are crucial for enabling multi-user beamforming with distributed antenna arrays.
We have provided experimental datasets to evaluate the impact of CFO in real-world systems, available at~\cite{dist_mu_mimo_datasets} for further research.
In our distributed scenario, while we refer to the placement of APs in Fig.~\ref{fig:exp_setup_ap} as distributed, they are only 3 meters apart.
This setup may not fully capture the characteristics of a truly distributed system with diverse channel conditions.
The main difference from a centralized case lies in the correlation pattern between antennas.
However, our focus is not on the effects of varying channel conditions, diversity patterns, or scattering scenarios.
Instead, we concentrate on the impact of CFO between distributed APs.
In this context, the dAPs may be physically close or have correlated channels, but they operate with independent LOs.
Thus, the word distributed in our study concentrates mainly on the use of independent RF frequency sources, potentially resulting in random frequency offsets between distributed nodes.

In our experiments in the LO-mode, we let every dAP use its own LO for deriving carrier and sampling frequencies, and there is no additional OTA frequency synchronization step performed.
However, in practical systems, there is always a frequency synchronization mechanism to ensure phase coherency among radios participating in communication.
We performed our experiments with free-running LOs to evaluate the performance in the worst-case scenario.
An extra synchronization step can compensate for the CFO-induced residual phase offset for some duration of time, depending on the synchronization algorithm.
To maintain phase coherence over time, the synchronization must be performed frequently, introducing communication overhead on the system that scales with the network size.
Moreover, the CFO-induced residual phase offset accumulates in distributed MU-MIMO systems as the signal is affected by multiple CFO levels from all dAP participants.
Several solutions and algorithms have been proposed in the literature for synchronization in distributed systems~\cite{gao2019_syncAlg, wang_2015_syncAlg, ahmad_2024_syncAlg, milos_2023_syncAlg}, some of which have shown effectiveness in enabling D-MUBF on a small scale~\cite{chorus, rogalin, rashid_2022_syncAlg, Rashid_2023_syncAlg}. Nevertheless, this work provides valuable insight into the impact of CFO on beamforming performance in distributed multi-user systems operating without perfect frequency synchronization.


\section{Conclusion}
\label{sec:conclusion}

In this work, we have conducted an analytical and experimental assessment of distributed multi-user beamforming in the presence of residual frequency offsets, addressing a crucial aspect of large-scale distributed MIMO systems.
Through an analytical framework, we introduced conditional closed-form expressions for SINR within a distributed MU-MIMO architecture.
Moreover, we conducted extensive experiments across various experiment scenarios using the RENEW massive MIMO testbed.
The datasets can be used to explore the impact of MU-MIMO configurations, distance between UEs, distance between dAPs, channel correlations, residual CFO from independent LOs, the number of beamforming antennas, as well as the beamforming scheme on the performance of distributed multi-user beamforming techniques.
In addition, we provided analytical, simulation, and experimental results to validate our analytical expressions and gain insights from the empirical measurements.
Our findings provide valuable insights into the impact of CFO on D-MUBF, facilitating the systematic design of parameters for distributed MIMO systems optimized for multi-user beamforming applications.

\appendices

\section{Derivation of Frequency Domain OFDM Signal}
\label{appendix_a}

In the presented OFDM signal model, the time-domain received signal at antenna $i$ in dAP $m$ prior to applying FFT is given in~\eqref{eq:ofdm_time_domain_rx_signal}.
By applying FFT on it to achieve the frequency domain representation, we obtain the following
\begin{align*}
    \label{eq:ofdm_fft_appendix}
    \mathrm{y}_m^{(i)}[n,l] & \textstyle = \frac{1}{\sqrt{N_{sc}}} \sum_{\zeta=0}^{N_{sc}-1} \mathrm{\check{y}}_m^{(i)}[n,\zeta]  e^{-j \frac{2 \pi}{N_{sc}} \zeta l} \\
    & \textstyle = \frac{1}{N_{sc}} \sum_{\zeta=0}^{N_{sc}-1} e^{j \frac{2 \pi}{N_{sc}} (\zeta+n(N_{sc}+L_{cp})) \epsilon_m} \cdot \\ 
    & \textstyle \quad \sum_{q=0}^{N_{sc}-1}  \mathrm{s}_k[n,q]  \mathrm{h}_{km}^{(i)}[q]  e^{j\frac{2\pi \zeta}{N_{sc}} (q-l)} \\
    & \textstyle = \frac{e^{j\overline{\phi}_{m}[n]}}{N_{sc}} \sum\limits_{q=0}^{N_{sc}-1} \sum\limits_{\zeta=0}^{N_{sc}-1} \mathrm{s}_k[n,q]  \mathrm{h}_{km}^{(i)}[q]  e^{j\frac{2\pi \zeta}{N_{sc}} (\epsilon_m + q-l)} \\
    & \textstyle = e^{j\overline{\phi}_{m}[n]} \sum_{q=0}^{N_{sc}-1} \mathrm{s}_k[n,q]  \mathrm{h}_{km}^{(i)}[q]  G_{\epsilon}[q-l] \\
    & \textstyle = e^{j\overline{\phi}_{m}[n]}  \mathrm{s}[n,l]  \mathrm{h}_{km}^{(i)}[l]  G_{\epsilon}[0] \\
    & \textstyle \quad + e^{j\overline{\phi}_{m}[n]}  \sum_{q\neq l}^{N_{sc}-1} \mathrm{s}[n,q]  \mathrm{h}_{km}^{(i)}[q]  G_{\epsilon}[q-l].
    \numberthis
\end{align*}

\section{Derivation of Conditional Powers}
\label{appendix_b}

To compute the conditional SINR based on~\eqref{eq:cond_sinr_main}, we require the conditional powers of the desired signal, interference, and noise. The conditional power of the desired signal, denoted by $P_{s|\mathbf{H}^{\mathrm{ul}}}$, is given by
\begin{align*}
    \label{eq:sig_cond_power_appendix}
    P_{s|\mbf{\mbf{H}^{\mrm{ul}}}}
    & \textstyle =\E\{| \sum_{m=1}^{M} \Omega_m \mathbf{w}_{km}^{\mathrm{H}} \mathbf{h}_{km} \mrm{s}_k |^2  |  \mbf{H}^{\mrm{ul}}\} \\
    & \textstyle = p_{\mrm{ul}} \sum\limits_{m=1}^M \sum\limits_{\mp =1}^M \E\{\Omega_m^* \Omega_{\mp}\} \cdot \\ 
    &\textstyle \qquad \E\{(\mathbf{w}_{km}^{\mathrm{H}} \mathbf{h}_{km} \mathbf{h}_{k\mp}^{\mrm{H}} \mathbf{w}_{k\mp})^* |  \mbf{H}^{\mrm{ul}} \} \\
    & \textstyle = p_{\mrm{ul}}  \sum\limits_{m=1}^{M} \E\{|\Omega|^2\} |\mathbf{w}_{km}^{\mathrm{H}} \mathbf{h}_{km}|^2 \numberthis \\
    & \textstyle \qquad + p_{\mrm{ul}}  \sum\limits_{m=1}^M \sum\limits_{\mp \neq m}^M \left(\E\{\Omega\}\right)^2  \left(\mbf{w}_{km}^{\mrm{H}} \mbf{h}_{km} \mbf{h}_{k\mp}^{\mrm{H}} \mbf{w}_{k\mp}\right).
\end{align*}
Similarly, the conditional power of the inter-user interference term, $P_{v|\mbf{\mbf{H}^{\mrm{ul}}}}$, can be computed as
\begin{align*}
    \label{eq:int_cond_power_appendix}
    P_{v|\mbf{\mbf{H}^{\mrm{ul}}}} &= \textstyle \E\{| \sum_{m=1}^{M} \Omega_m \sum_{u\neq k}^K \mathbf{w}_{km}^{\mathrm{H}} \mathbf{h}_{um} \mrm{s}_u |^2  |  \mbf{H}^{\mrm{ul}}\} \\
    & \textstyle = \sum\limits_{m=1}^M \sum\limits_{\mp =1}^M \E\{\Omega_m^* \Omega_{\mp}\} \cdot \\
    & \qquad \textstyle \E\{\sum_{u\neq k}^K \mathbf{w}_{km}^{\mrm{H}} \mathbf{h}_{um} \mrm{s}_u \sum_{u^\prime \neq k}^K \mathbf{w}_{k \mp}^{\top} \mathbf{h}_{u^\prime \mp}^* \mrm{s}_{u^\prime}^*\} \\
    &\textstyle = p_{\mrm{ul}}  \sum\limits_{m=1}^{M} \E\{|\Omega|^2\} \sum\limits_{u\neq k}^K \sum\limits_{\up \neq k}^K \mathbf{w}_{km}^{\mathrm{H}} \mathbf{h}_{um} \mathbf{h}_{\up m}^{\mathrm{H}} \mathbf{w}_{km} \\
    & \qquad \textstyle + p_{\mrm{ul}} \sum\limits_{m=1}^M \sum\limits_{\mp \neq m} \left(\E\{\Omega\}\right)^2 \cdot \\
    & \qquad \textstyle \sum\limits_{u\neq k}^K \sum\limits_{\up \neq k}^K \mathbf{w}_{km}^{\mathrm{H}} \mathbf{h}_{um} \mathbf{h}_{\up \mp}^{\mathrm{H}} \mathbf{w}_{k\mp}.
    \numberthis
\end{align*}




\bibliographystyle{IEEEtran}
\bibliography{bibtex/bib/IEEEabrv, bibtex/bib/ref}

%

\begin{IEEEbiography}[{\includegraphics[width=1in,height=1.25in,clip,keepaspectratio]{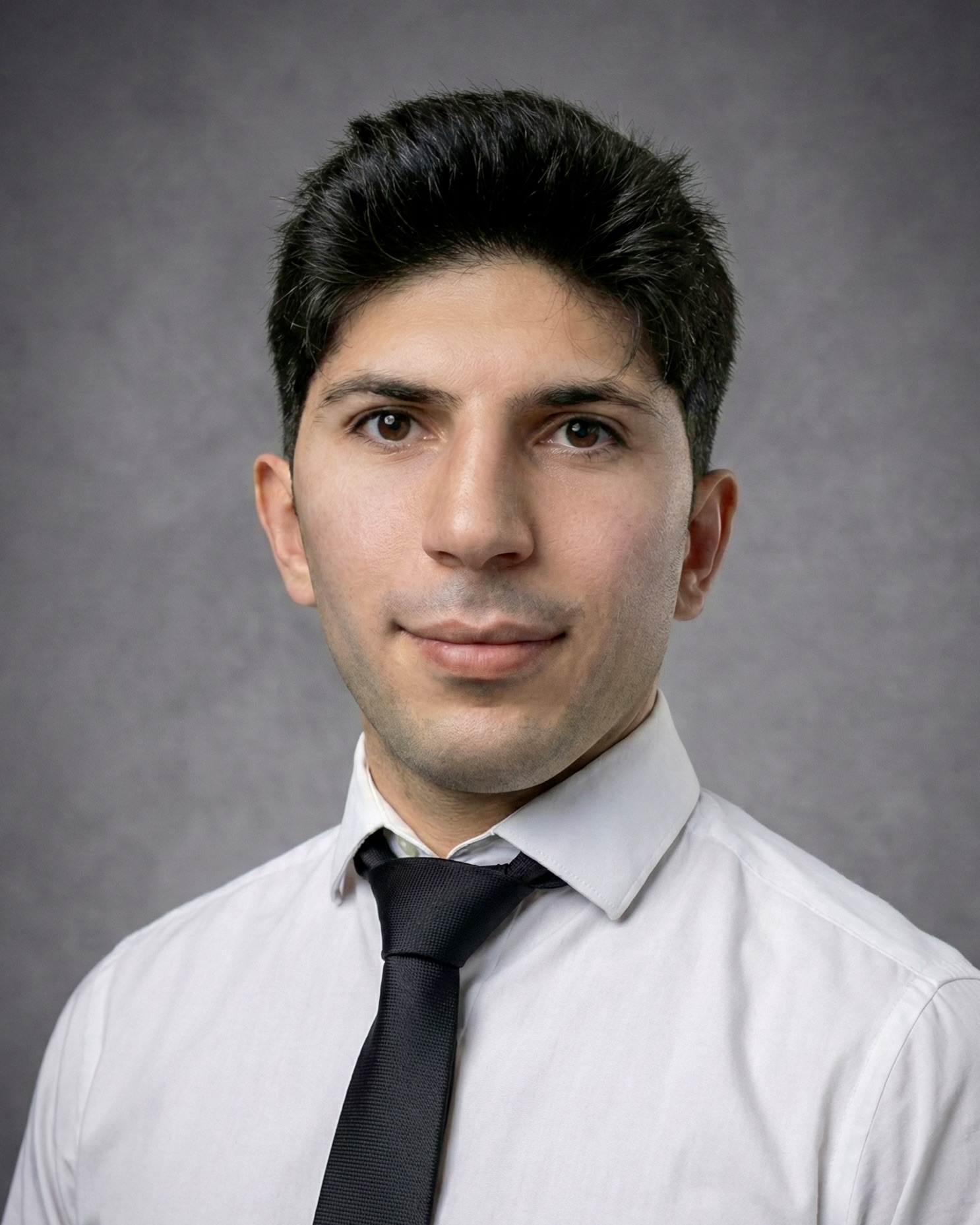}}]{Mehdi Zafari}
(S'21) received the B.S. degree in Electrical Engineering from the University of Tehran, Tehran, Iran, in 2015, and the M.S. degree in Electrical and Computer Engineering from Rice University, Houston, TX, USA, in 2024. He is currently pursuing the Ph.D. degree in Electrical Engineering and Computer Science at the University of California, Irvine, CA, USA.
His research interests include optimization, deep learning, and generative AI for next-generation wireless communications, with a focus on integrated sensing and communication (ISAC), cell-free/distributed MIMO technology, and resource allocation.
\end{IEEEbiography}

\begin{IEEEbiography}[{\includegraphics[width=1in,height=1.25in,clip,keepaspectratio]{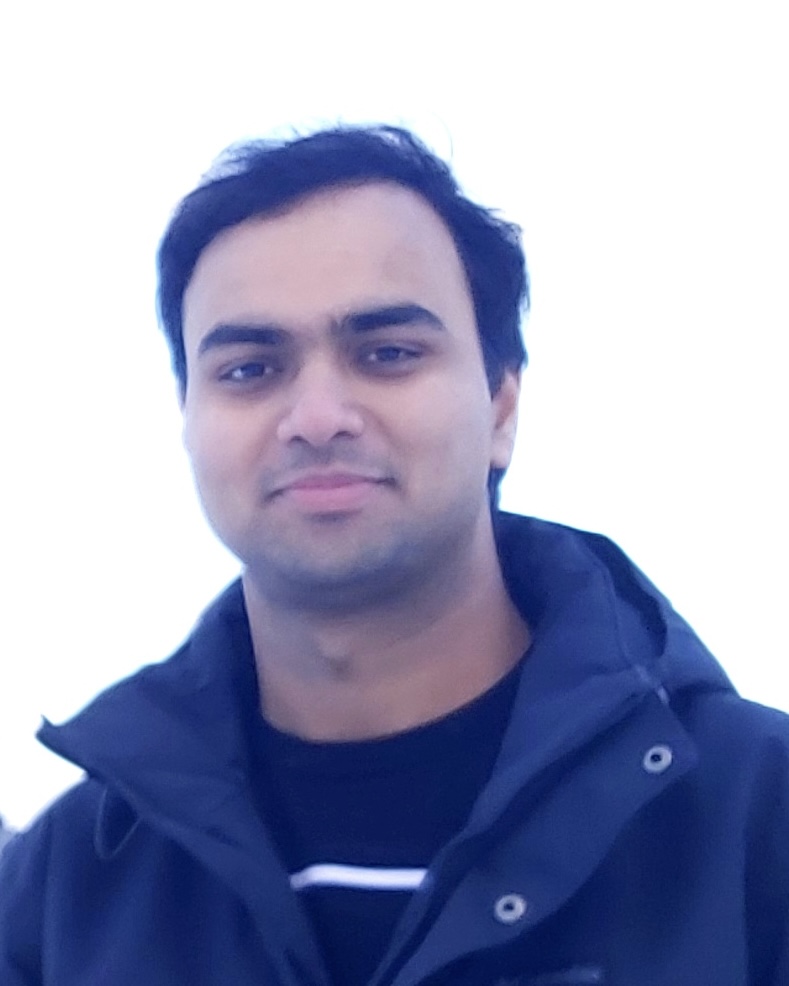}}]{Divyanshu Pandey}
received the B.Tech. degree in Communication and Computer Engineering from the LNM Institute of Information Technology, Jaipur, India, in 2011, the M.S. degree in Electrical Engineering from the University of Minnesota, Twin Cities, MN, USA, in 2014, and the Ph.D. degree in Electrical Engineering from McGill University, Montreal, QC, Canada, in 2022.

From 2011 to 2013, he was an Instrumentation Engineer with HMEL, Bathinda, India. He was a Wireless Systems Engineer with Marvell Semiconductor, Santa Clara, CA, USA, from 2015 to 2017, and a Postdoctoral Associate with the Department of Electrical and Computer Engineering at Rice University, Houston, TX, USA, from 2023 to 2025. He is currently working as a Wireless Systems Engineer with Apple, Sunnyvale, CA, USA.

His research interests include wireless communication systems, information theory, joint sensing and communication systems, and tensor algebra with applications to communications and signal processing.

\end{IEEEbiography}

\begin{IEEEbiography}[{\includegraphics[width=1in,height=1.25in,clip,keepaspectratio]{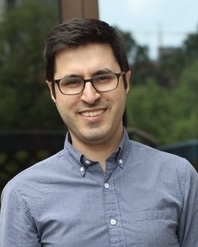}}]{Rahman Doost-Mohammady}
received his Ph.D. degree from Northeastern University, Boston, MA, USA, in 2015. He is currently an assistant research professor in Electrical and Computer Engineering at Rice University, Houston, TX, USA. His current research interests involve programmable and open radio access networks, applied machine learning in wireless systems, and the development of large-scale experimental next-generation wireless platforms.

\end{IEEEbiography}

\vfill







\end{document}